# PROPERTIES OF A COMPACT NEUTRON SUPERMIRROR TRANSMISSION POLARIZER WITH AN ELECTROMAGNETIC SYSTEM

V.G. Syromyatnikov[1,2], S.Yu. Semenikhin[1], M.V. Lasitsa[1]

1 - *Petersburg Nuclear Physics Institute (PNPI), National Research Center "Kurchatov Institute", Gatchina, Russia*
2 - *Department of Physics, Saint Petersburg State University, St. Petersburg, Russia*

E-mail address: syromyatnikov_vg@pnpi.nrcki.ru

## Abstract

The paper will present *SVAROG*, a compact neutron supermirror transmission multichannel polarizer with an electromagnetic system. The basic properties of this polarizer will be considered. Variants for using this polarizer in experimental facilities of the *PIK* research reactor (Petersburg Nuclear Physics Institute of National Research Centre «Kurchatov Institute» (NRC «Kurchatov Institute» - PNPI)) will be discussed and a comparison of the considered polarizer with known neutron transmission polarizers will be carried out.

## 1. Introduction

In [1, 2], a neutron transmission supermirror kink polarizer of a new type was proposed and briefly considered. In [3, 4], a proposal was considered to increase the luminous intensity of this polarizer by adding a second element, a direct polarizing neutron guide. At the same time, the angular range of the outgoing beam with high polarization has increased significantly, several times. This polarizer is designed to operate in small magnetic fields, in which the remanent properties of polarizing supermirrors can be used.

In [5], a polarizer was considered in which the elements (kink and direct polarizing neutron guide) are located in saturating magnetic fields. In addition, a spin-flipper has been added between these elements. It is shown that the main characteristics of this polarizer are high and it can be used for a number of neutron physics facilities of the new PIK research reactor (NRC «Kurchatov Institute» - PNPI).

In this paper, as part of the development of the above-mentioned papers, *SVAROG*, a compact neutron transmission supermirror polarizer with an electromagnetic system, will be presented. The use of this electromagnetic system makes it possible to optimize the polarizer design and reduce its length. The main characteristics of the polarizer, depending on the parameters of its elements, have been obtained and discussed in detail. The variants for using the *SVAROG* polarizer in neutron physics facilities of the *PIK* reactor are considered. This polarizer is compared with known transmission neutron polarizers.

The material of this paper was presented as an invited report at the Russian Conference on the Use of Neutron Scattering in Condensed Matter Research (RNIKS-2025) [6].

The contents of the sections of this paper are: 1st section – Introduction; 2nd section – Description of the properties of neutron polarizing *CoFe/TiZr* supermirrors; 3rd section – Description of new type transmission polarizer variants; 4th section – Description of the proposed *SVAROG* transmission polarizer variant including the electromagnetic system; 5th section - Description the calculations of the intensity of both spin components of the beam passing through *SVAROG*; the 6th section - Description the calculations of the spectral dependences of polarization and beam transmission for a *SVAROG* polarizer with air channels, taking into account absorption and scattering in silicon; 7th section – Comparison of the *SVAROG* polarizer with analogues; Conclusions; Statement of the author's contribution; Statement of competing interests; Acknowledgements; Appendix A; Appendix B; References.

## 2. Description of the properties of neutron polarizing *CoFe/TiZr* supermirrors.

### 2.1. Neutron polarizing *CoFe/TiZr* ($m$ = 2.0; 2.5) supermirrors in saturating magnetic fields.

Figures 1a, b show, the reflectivities curves $R^+$ and $R^-$ for both spin components of the beam for highly efficient neutron *CoFe/TiZr* ($m$ = 2.0 [7, 8] and 2.5 [9]) supermirrors developed at the PNPI. The supermirrors are located in a saturating magnetic field. As follows from the figures, for the (+) spin component of the beam, when the neutron spin is parallel to the magnetization vector of the magnetic layers of the supermirrors, the reflection coefficient $R^+$ is high, because the neutron-optical potential of the magnetic layer for this component significantly exceeds the potential of the non-magnetic layer.

The critical beam reflection angle from the supermirror is also high. For the (-) spin component of the beam, when the neutron spin is antiparallel to the magnetization vector of the magnetic layers of the supermirror, the reflection coefficient $R^-$ is negligible, since the neutron-optical potential of the magnetic layer for this component is small and is practically equal to the potential of the non-magnetic layer for the saturating value of the magnetic field. Accordingly, the critical beam reflection angle from the supermirror for this component is very small. Reflection from the substrate is insignificant, because an antireflective absorbing *TiZrGd* layer is sputtered between the substrate and the supermirror.

The graphs shown in Fig. 1 will be used in the calculations presented in this paper.

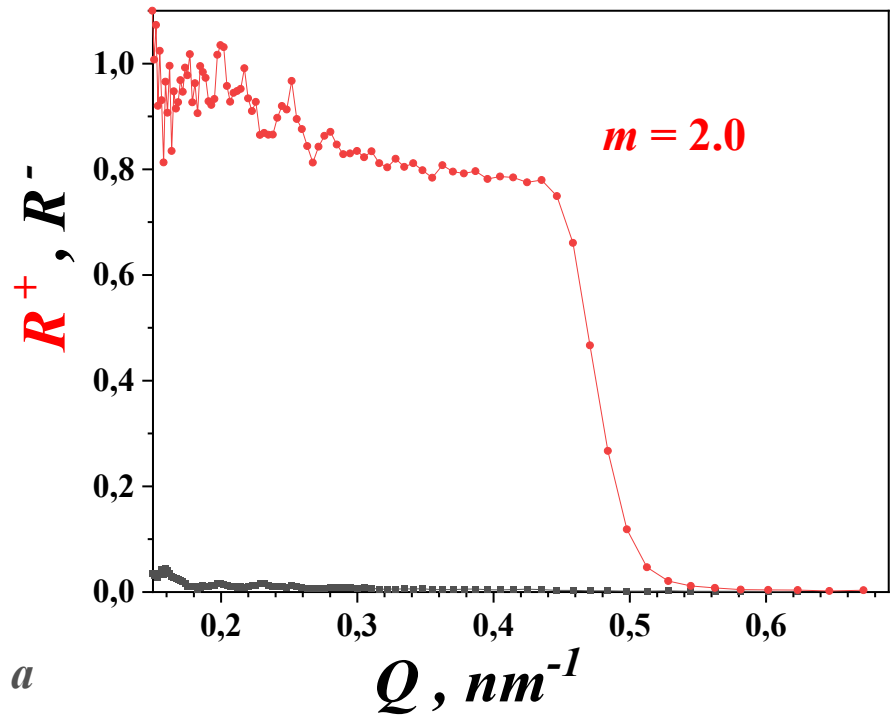

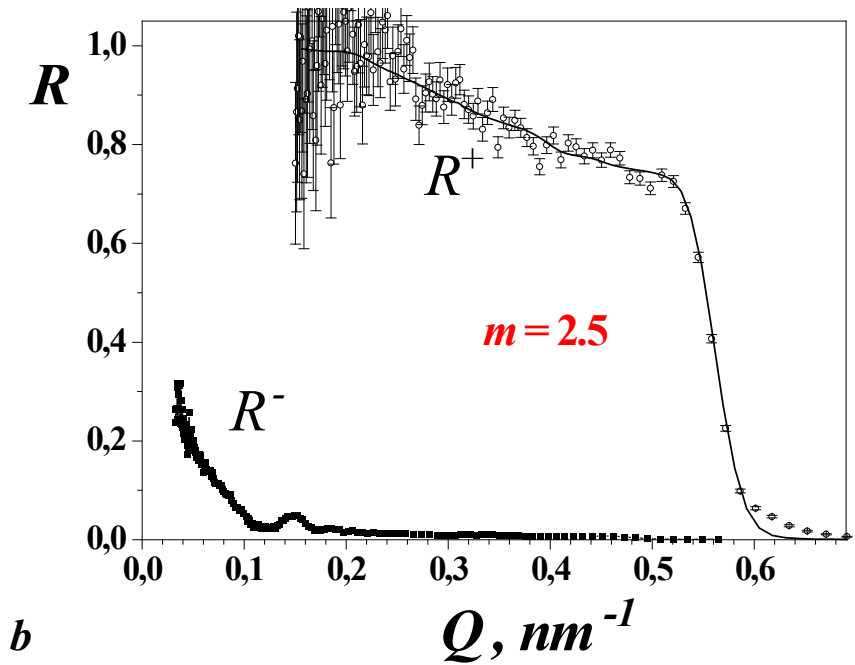

**Fig. 1a, b.** The reflectivities curves $R^+$ and $R^-$ for both spin components of the beam for highly efficient neutron *CoFe/TiZr* ($m = 2.0$ (a) and 2.5 (b)) supermirrors developed at the PNPI. The supermirrors are located in a saturating magnetic field.

## 2.2. Remanent properties of polarizing *CoFe/TiZr* supermirrors.

As is known, the magnetron sputtering technique is widely used to create polarizing neutron periodic and aperiodic (supermirrors) multilayer structures. During this deposition, magnetic anisotropy with light and hard magnetization axes occurs in the magnetic layers of the supermirror coating. The magnetization curve of such a multilayer structure (supermirror) along the light axis is characterized by high **_remanence_**, i.e. in the region of small fields of ~ (10-20) Gs on both branches of the hysteresis curve, the magnetization of the magnetic layers is high and close to its maximum value in absolute terms. The shape of the hysteresis loop is close to a rectangle, as shown in Fig. 2 [4]. If we are at point *A* with $H = H_r$ of the hysteresis curve, the magnetization of the magnetic layers of the multilayer structure is oriented parallel to the guiding field and the reflected beam will be polarized along the field (see Fig. 2). If we are at point *B* of the hysteresis curve, with the same magnitude of the $H_r$ field, the magnetization of the magnetic layers of the supermirror is oriented antiparallel to the guiding field and the reflected beam will be polarized antiparallel to the field (see Fig. 2).

Using the remanence property of polarizing supermirrors allows you to work in a polarization analysis scheme without a spin-flipper in front of the sample. A number of works are known in which the remanent properties of periodic and aperiodic (supermirrors) multilayer structures of *CoFe/TiZr* and *Fe/Si* have been described [8, 10, 11].

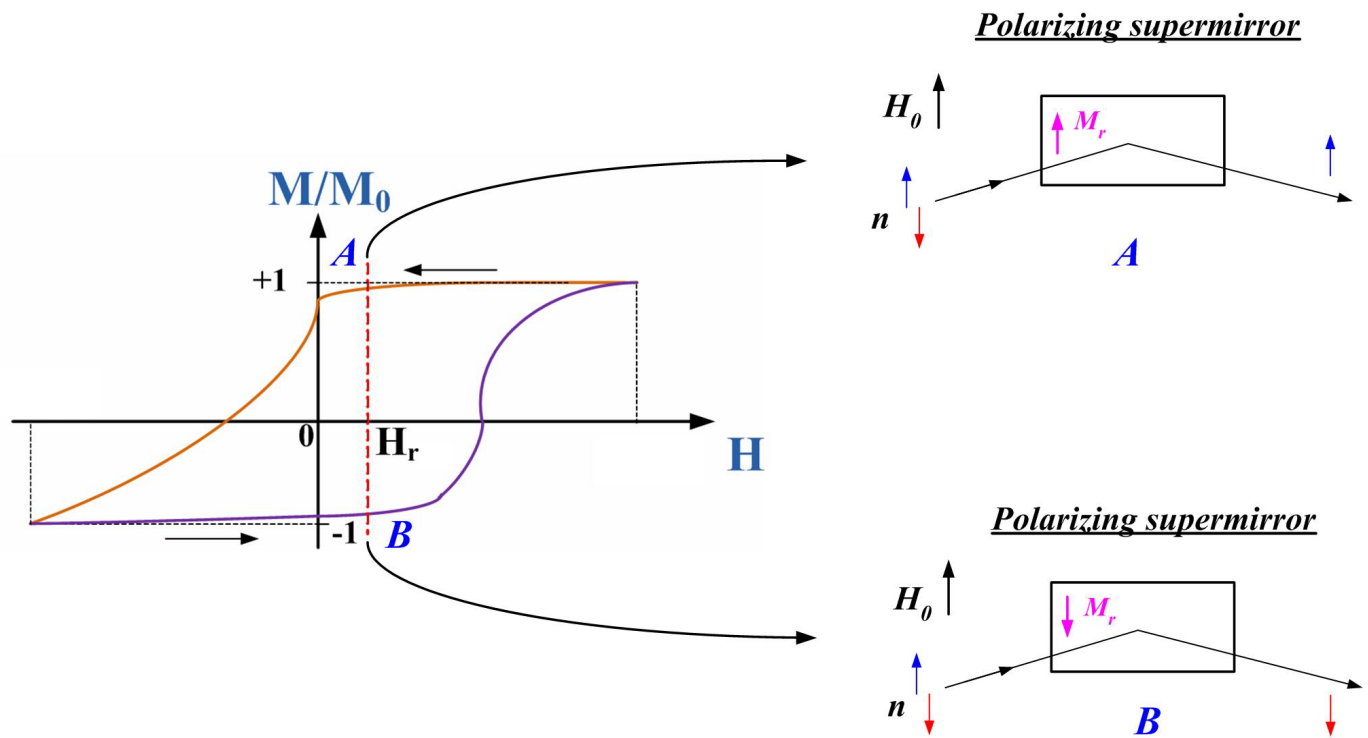

**Fig. 2.** Hysteresis loop for neutron remanent polarizing supermirrors.

As an example, Figure 3 shows the reflectivities curves $R^+$ and $R^-$ for both spin components of the beam for the remanent *CoFe/TiZr* (*m* = 2) supermirror No. 104 for the saturating value of the magnetic field [4]. Figure 4 shows the dependence of the integral polarizing efficiency on the remagnetization of this supermirror. As follows from the graph, supermirror No. 104 has pronounced remanent properties. The graphs shown in Figs. 3 and 4 are obtained from measurements carried out on the *NR-4M* (PNPI) neutron reflectometer [12].

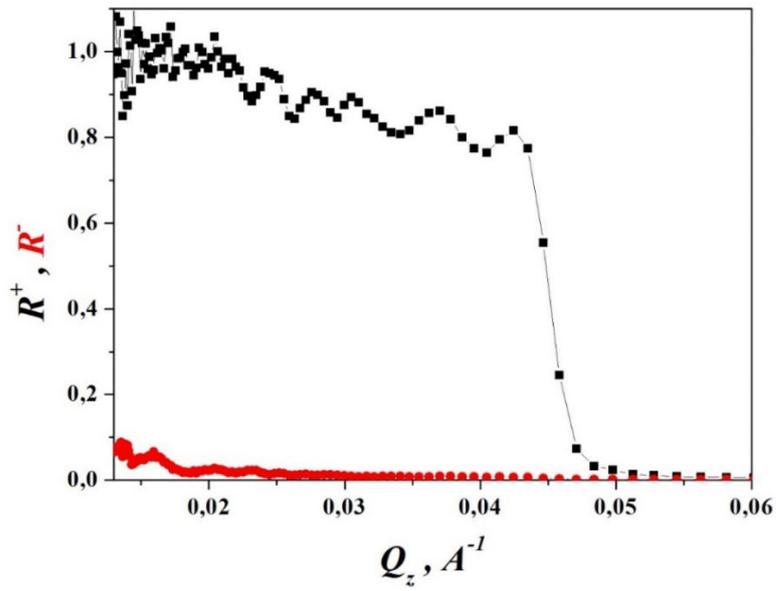

**Fig. 3.** The reflectivities curves $R^+$ and $R^-$ for both spin components of the beam for the remanent $CoFe/TiZr$ ($m = 2$) supermirror No. 104 for the saturating value of the magnetic field [4].

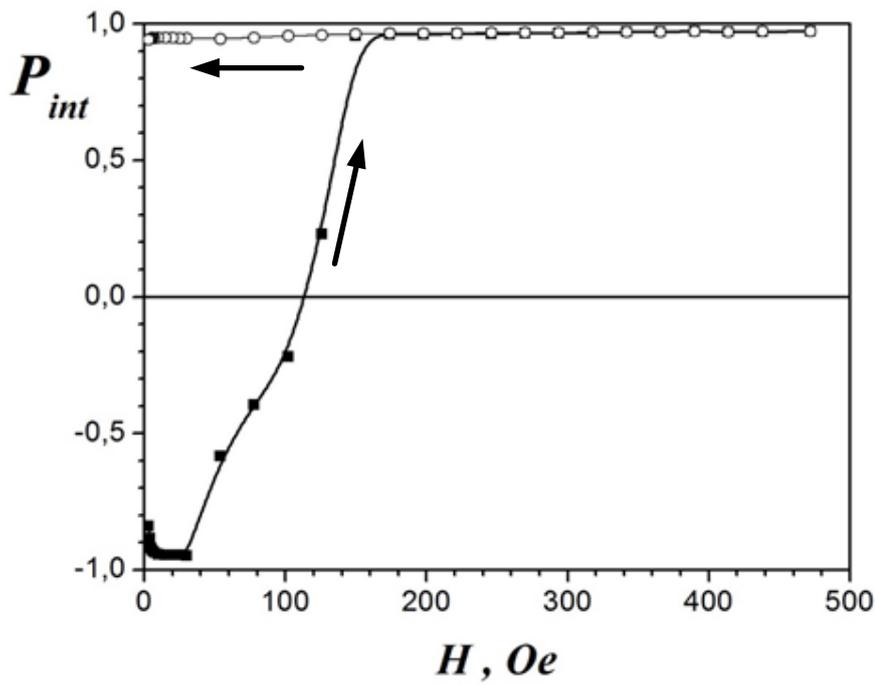

**Fig. 4.** Dependence of the integral polarizing efficiency of supermirror No. 104 during its remagnetization.

# 3. Description of new type transmission polarizer variants.

## 3.1. Transmission kink polarizer.

Figure 5 shows a scheme of a kink polarizer consisting of two separate parts (shoulders) [5]. Each part is a stack of $N$ identical silicon wafers pressed together without air gaps. All the wafers are polished on both sides. The lengths of the wafers are $L_1$ and $L_2$, respectively. The stacks are tilted relative to the horizontal axis at angles $\theta_1$ and $\theta_2$, respectively. The wafer thicknesses are the same and equal to $d$. A polarizing *CoFe/TiZr* supermirror coating with parameter $m = 2$ without an absorbing *TiZrGd* layer is sputtered to each side of the wafer. The kink's parameters are shown in Table 1. To calculate the transmission of a neutron flux through such a kink, a program prepared in the *McStas* platform was used in accordance with the scheme in Fig. 5 and the parameters from Table 1.

**Table 1.** Parameters of the neutron kink polarizer.

| $D$, mm | $L_1$, mm | $L_2$, mm | $\theta_1$, mrad | $\theta_2$, mrad | $m$ |
|---|---|---|---|---|---|
| 0.30 | 41 | 64 | 7.4 | 4.7 | 2 |

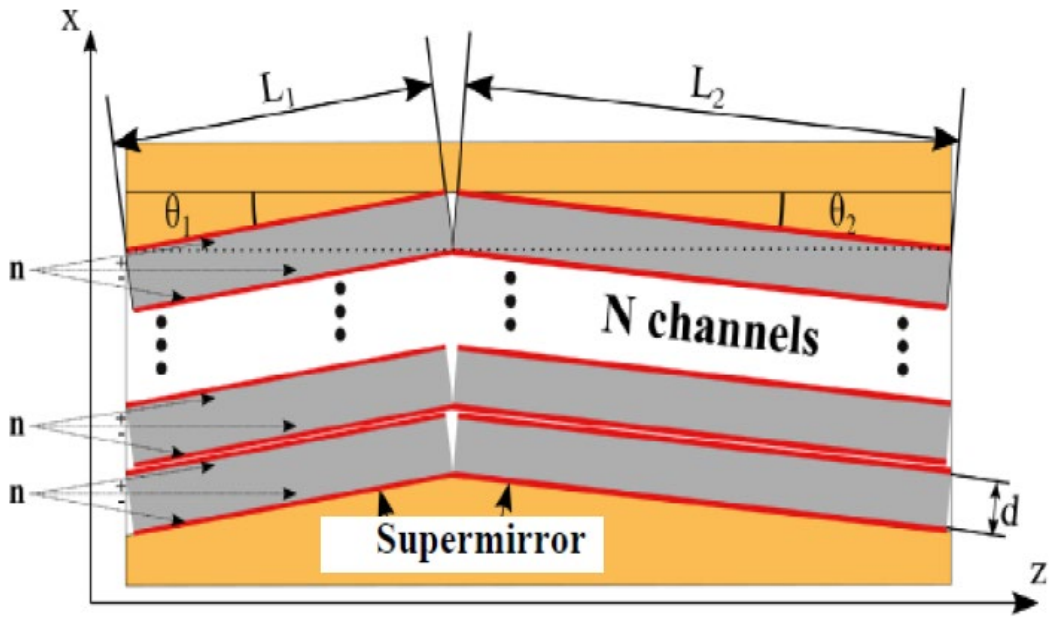

**Fig. 5.** The scheme of a kink polarizer consisting of two separate parts (shoulders).

As noted above, for these supermirrors, the neutron-optical potentials of the supermirror layers for the (-) spin components of the beam are almost equal to each other for the saturating magnetic field applied to the supermirror in its plane, and these potentials are close to the substrate potential and do not exceed it. Therefore, the critical angle for the "material-supermirror" boundary is close to zero. In this regard, the reflection coefficient of this component of the beam from the supermirror is minimal, i.e. the neutrons of this component practically do not deviate from their initial trajectories when passing through the kink. At the same time, for the (+) component of the beam, the potential of the magnetic layer is significantly higher than the potentials of the non-magnetic layer and the substrate. This makes it possible to obtain a high neutron reflection coefficient of this component from the supermirror and a high critical angle for the "material-supermirror" boundary. Consequently, the neutrons of this component of the beam will deviate significantly from their original trajectories when passing through the kink.

As a polarizing supermirror coating in kink, a polarizing *Fe/Si* supermirror can also be used, for which the same conditions are met as for *CoFe/TiZr* supermirrors.

To illustrate the above, Figure 6 shows a scheme of the trajectories of the neutrons (+) and (-) spin components of the beam as they pass through the kink polarizer. It follows from this scheme that the width of the angular distribution at the output of the polarizer for the (+) spin component of the beam significantly exceeds the same value for the (-) component, i.e. $\alpha_{up} > \alpha_{down}$.

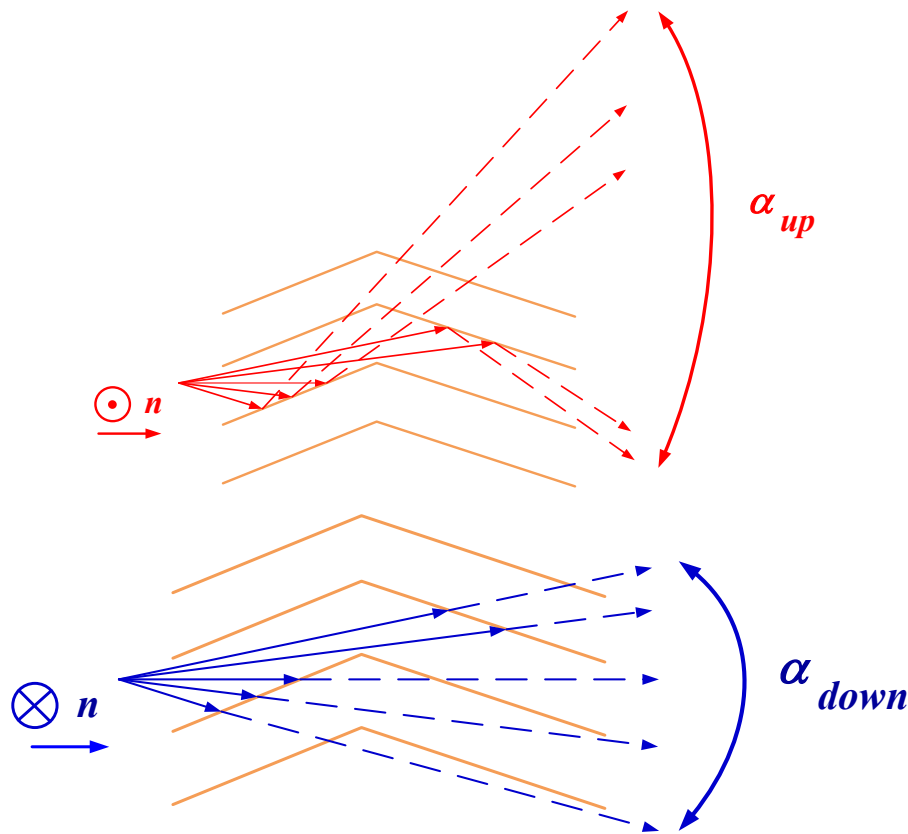

**Fig. 6.** The scheme of the trajectories of the neutrons of the (+) and (-) spin components of the beam as they pass through the kink polarizer.

Figure 7 shows the results of calculations of the transmission of a neutron beam with a wavelength of 4 Å through a kink polarizer, depending on the angle for both spin components of the beam [5]. The parameters of the kink polarizer are shown in Table 1. The black line shows the angular distribution of the beam intensity at the entrance to the polarizer with a divergence of ± 0.45 degrees and the angular distribution of the intensity of the (-) component at the output of the kink. The red color shows the curve of the angular intensity distribution at the kink output for the (+) spin component. The angular intensity distribution of the (+) component at the kink output contains a set of noticeable peaks caused by the reflection of neutrons of this component from supermirror coatings. At the same time, there is a dip in intensity for this component near the axis of the incident beam. The angular intensity distribution of the (-) component at the kink output, as follows from the figure, coincides with the input angular distribution, since the absorption and scattering of neutrons in silicon were not taken into account here and the reflection of neutrons of the (-) component from supermirror coatings was neglected.

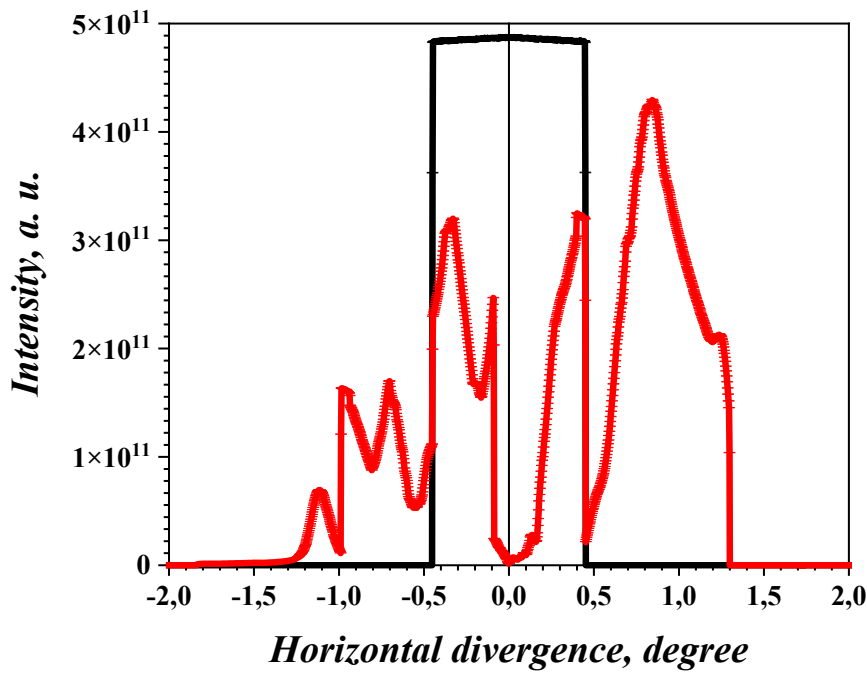

**Fig. 7.** Calculated transmission of the intensity of a neutron beam with a wavelength of 4 Å through a kink polarizer, depending on the angle for both spin components of the beam. Here, the intensity at the input to the polarizer and the (-) spin component at the output of the polarizer are shown in black, while the intensity of the (+) spin component at the output of the polarizer is shown in red.

Figure 8 shows the same curves as in Figure 7, but on a smaller scale. All peaks of the (+) spin component must be removed from the output beam. This will require installing a multichannel Soller collimator with "black" non-reflecting and neutron-absorbing walls at the output of the polarizer. This collimator must pass a beam with a divergence not exceeding ± 0.1 degrees (this angular range is shown in the blue dotted line in the figure) in order to pass the neutrons of the (-) spin component and remove all peaks of the (+) component from the beam. In Fig. 9a, b show a scheme of one channel of a Soller collimator on a silicon substrate (a) and the angular dependence of the transmission coefficient of the neutron flux through this device (b). Here $L3 = 172$ mm and $d = 0.3$ mm.

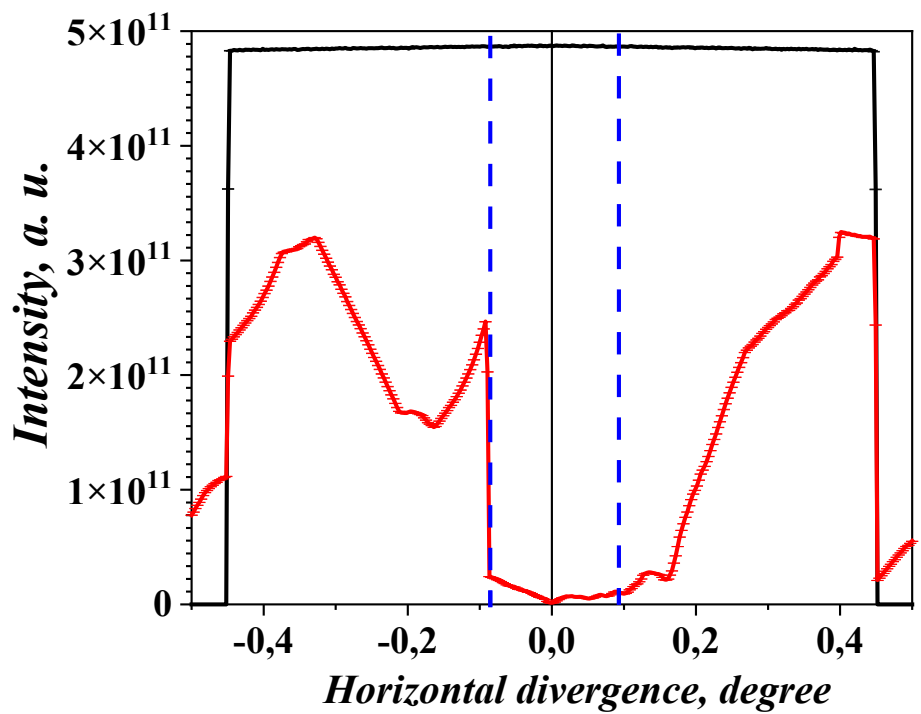

**Fig. 8.** Shows the same curves as in Figure 7, but in a smaller angular range. The blue dotted line highlights the angular range in which the polarization of the beam at the output of the polarizer is high.

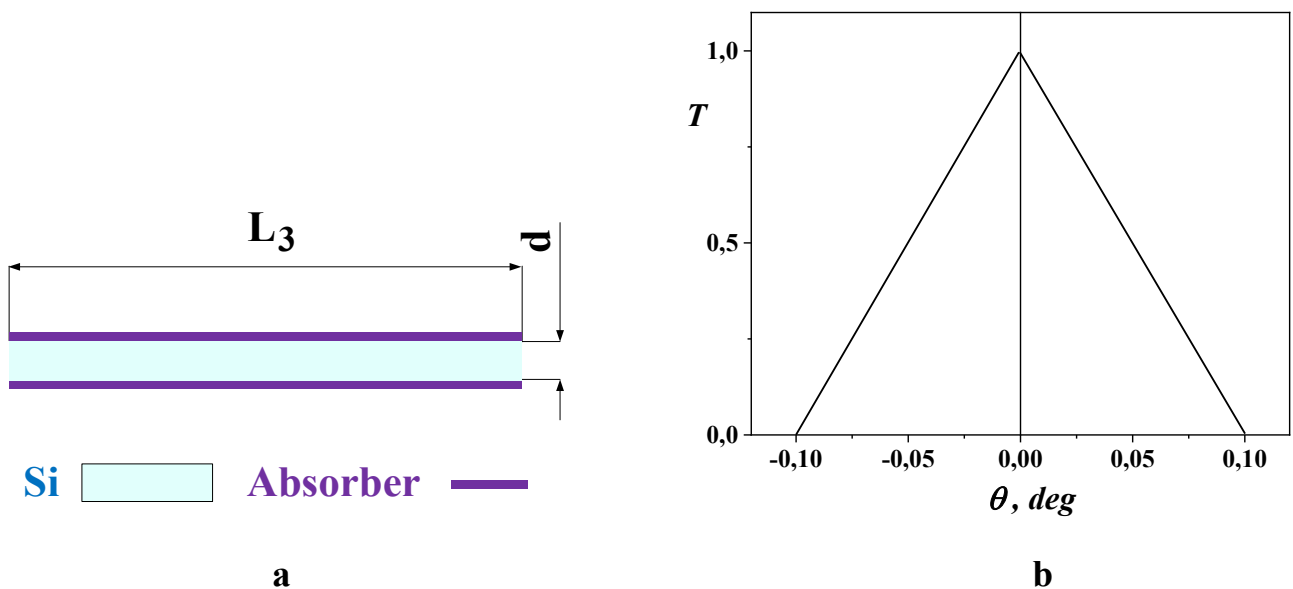

**Fig. 9a, b.** The scheme of a single channel of a Soller collimator on a silicon substrate (a) and the angular dependence of the transmission coefficient of the neutron flux through this device (b).

Thus, at the output of the kink polarizer, the beam will have a high level of negative polarization. Unfortunately, the angular width of this beam, as follows from Fig. 8, will be almost 5 times less than the beam width at the entrance, i.e. the beam transmission through the kink and the aperture will not be high enough.

### 3.2. Remanent supermirror polarizer

A compact neutron remanent transmission polarizer was proposed and considered in [3, 4]. Let's consider the concept of this polarizer in accordance with the work [4]. This is a supermirror solid state multichannel transmission polarizer. It uses the property of the remanence of polarizing supermirrors. The polarizer consists of two compact multichannel solid state parts. In both parts, each channel is a plate of material transparent to neutrons. The first part is the spin splitter, the second part is the straight polarizing neutron guide or collimator-polarizer (see Fig. 10). In the 1st part, a supermirror coating is applied directly to a substrate transparent to neutrons on both sides. In the 2nd part, a supermirror coating is also applied to both sides of the substrate. Then an anti-reflective absorbing sublayer is applied to it. The substrate also absorbs neutrons. A kink polarizer can be used as a spin splitter.

The potentials of the supermirror layers with magnetization $M_r$ for the spin component of the inverse of the $M_r$ vector are close to each other when the magnetic layers are in a remanent state. At the same time, these potentials are close to the potential of the substrate material and do not exceed its value. The parts of the polarizer are inversely magnetized with respect to each other, and one of them is parallel to the guiding magnetic field $H_0$ (see Fig. 10). Neutrons of both spin components leave the 1st part of the polarizer. Only the neutrons of the spin component of the beam, which is opposite to the guiding field, exit from the 2nd part of the polarizer.

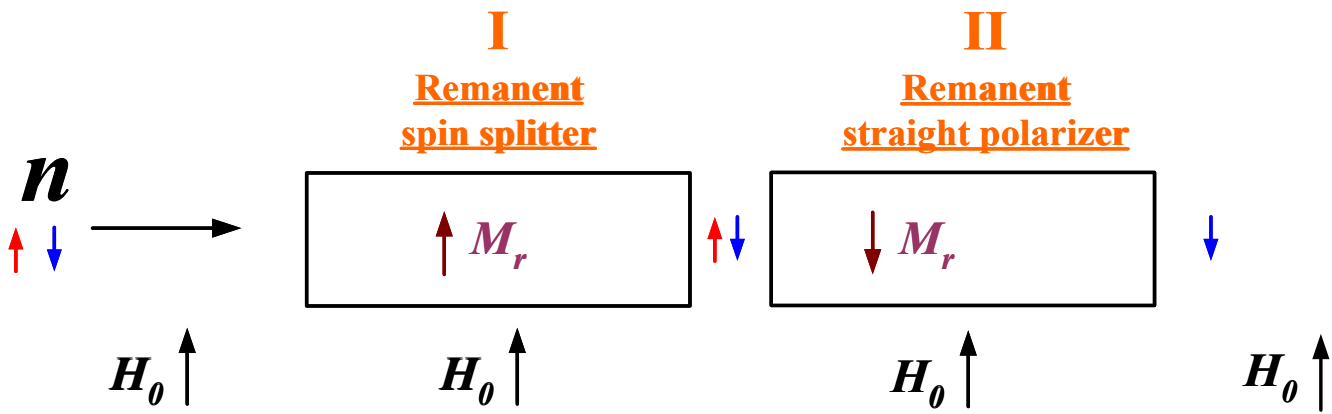

**Fig. 10.** The scheme of a neutron remanent supermirror polarizer [4].

The reflectivities curves $R^+$ and $R^-$ for both spin components of the beam for remanent polarizing supermirrors ($m = 2$) located in the 1st and 2nd parts of the polarizer are shown in (Fig. 11a) and (Fig. 11b), respectively. As follows from the figures, the curves are inverse to each other.

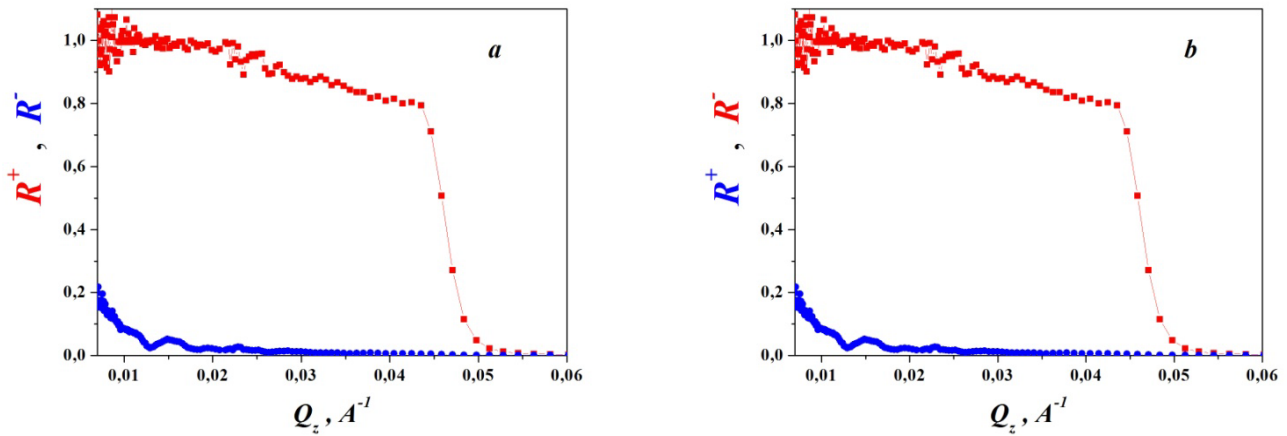

**Fig. 11a, b.** The reflectivities curves $R^+$ and $R^-$ for both spin components of the beam for remanent polarizing supermirrors ($m = 2$) located in the 1st (a) and 2nd (b) parts of the polarizer.

Figure 12 shows a scheme of a compact remanent polarizer (top view) consisting of a kink polarizer (1st part) and a straight polarizing neutron guide (2nd part), which are three stacks of silicon wafers pressed against each other without air gaps. The figure also shows the wafer schemes for the 1st and 2nd parts of the polarizer, respectively. The kink and the neutron guide have oppositely directed magnetizations of their supermirror coatings. The direction of the kink magnetization coincides with the direction of the guiding magnetic field. Its vector is perpendicular to the drawing plane.

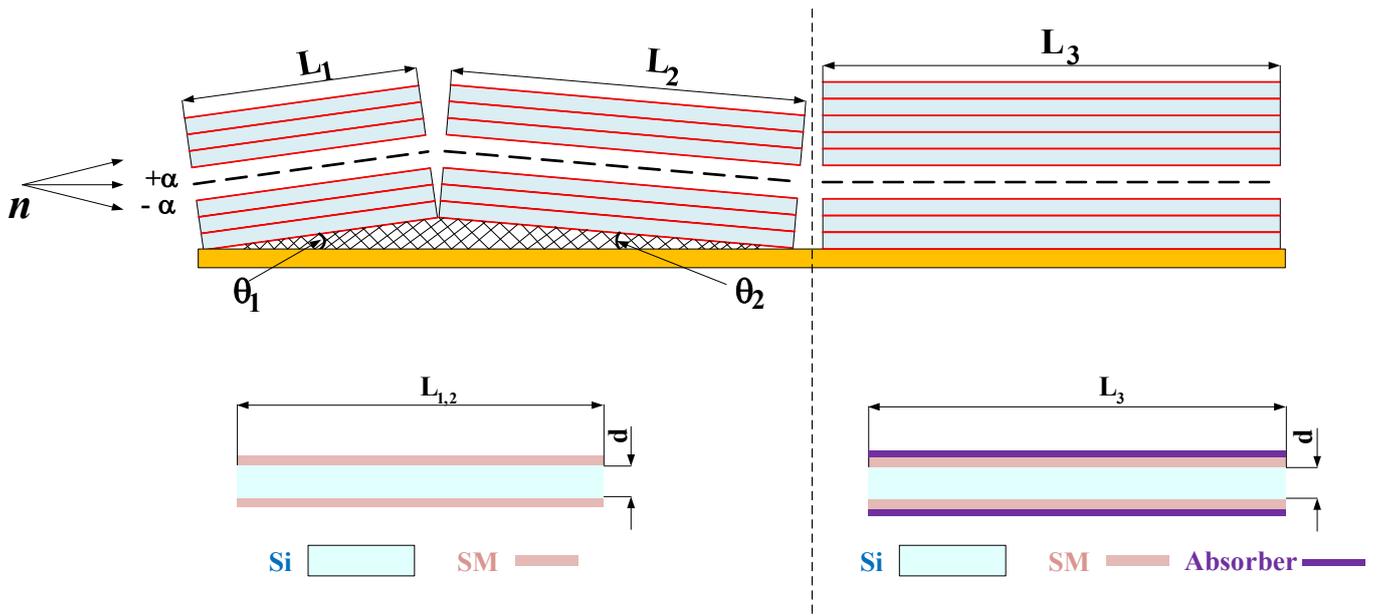

**Fig. 12.** A scheme of a compact remanent polarizer (top view), consisting of a kink polarizer and a straight polarizing neutron guide, representing three stacks of silicon wafers pressed against each other without air gaps. The figure also shows the wafer schemes for the 1st and 2nd parts of the polarizer, respectively.

Let us consider in detail the passage of each spin component of the beam through this polarizer using Figs. 13 and 14. The beam profile at the input is the same for both spin components.

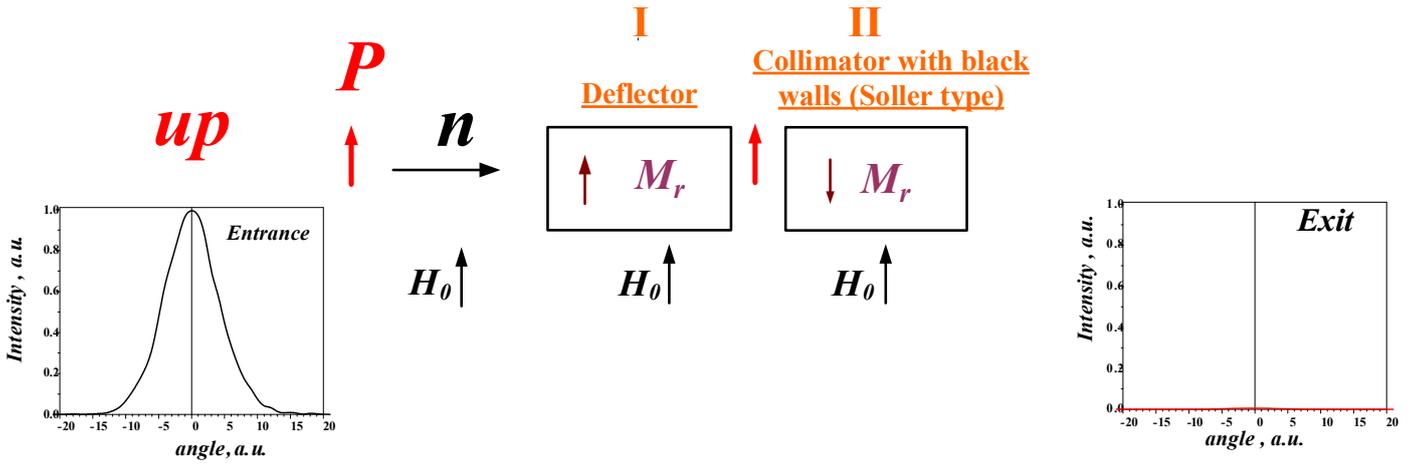

**Fig. 13.** Scheme of passage (+) spin component of the beam through the polarizer.

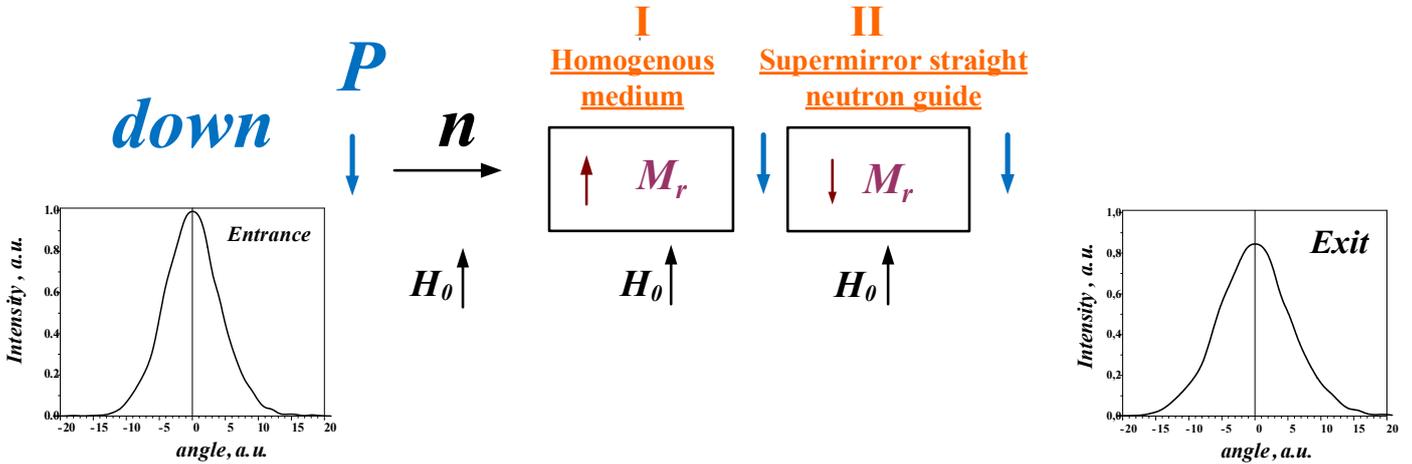

**Fig. 14.** Scheme of passage (-) spin component of the beam through the polarizer.

For (+) spin component of the beam (Fig. 13) the 1st part of the polarizer is a deflector, i.e. it significantly deflects the neutrons this component of the beam from their original trajectories due to the reflections of the neutrons of this spin component from the supermirror coatings in the channels of 1st part of the polarizer. In this case, the divergence of the beam increases significantly (see Fig. 6), and in the area of angles close to the beam axis falls to almost zero. When passing through the 2nd part of the polarizer, the neutrons of this spin component are reflected from the walls with a very low reflectivity, because the orientations of neutron spins and magnetization vectors of the magnetic layers of supermirrors are antiparallel.

The non-reflected neutrons of this spin component are absorbed in the sublayer of the supermirror. In fact, in this case, the 2nd part works as a conventional Soller collimator (see Fig. 9) with walls made from neutron-absorbing and non-reflecting material. It would seem that neutrons must pass through it, because it has straight visibility in the angular range given by the width and length of each of its channels. But there are no neutrons in this angular range, because they were deflected from their original trajectories in the 1st part of the polarizer. As a result, there are practically no neutrons (+) spin component at the exit of the polarizer. Thus, in the 1st part, the polarizer deflects the neutrons of this component, and then absorbs them in the 2nd part. Thus, for this spin component of the beam, the polarizer performs two functions: **deflect and absorb!**

For neutrons (-) spin component of the beam, the situation is different (see Fig. 14). For them, the 1st part of the polarizer is a homogeneous medium, which they pass without deviations from their original trajectories (see Fig. 6), only slightly reducing their intensity due to absorption in the substrate material.

This is due to the fact, as mentioned earlier, that the neutron-optical potentials of the supermirror layers and the substrate material are close for spin component of the beam which inverse to the guiding field. These neutrons pass through the 2nd part of the polarizer as through a conventional straight neutron guide, reflecting off its walls with a high reflection coefficient (Fig. 11b), because in this case, the neutron spins and the magnetization of the magnetic layers of the supermirror are parallel to each other. As is known, the beam does not change its divergence when passing through a straight neutron guide.

Thus, at the exit of the polarizer, the angular profile of the beam of this spin component repeats the beam profile at the entrance to the polarizer with a slight attenuation due to absorption in the substrate material (Fig. 14). As a result, the nonpolarized neutron beam passing through both parts of the polarizer will have a high negative polarization at the exit of the polarizer.

This compact transmission polarizer has high output beam intensity and polarization values when operating in small magnetic fields. However, there are certain difficulties in using it in the experiment. So, in order to create different magnetization in both parts of the polarizer, it is necessary to disassemble it and magnetize the neutron guide of the polarizer in a separate magnet. Then connect both parts again. It's not very convenient. This question needs to be specifically addressed. The solution to this problem will be discussed in section 4 of this paper.

**3.3. Transmission polarizer operating in saturating magnetic fields.**

In the paper [4], a polarizer scheme is proposed in which the kink and the neutron guide have independent magnetic systems on permanent magnetic elements that create high saturating magnetic fields. The directions of these fields coincide with each other, as well as with the direction of the guiding field. This is done in order to eliminate the inconveniences that arise when creating reverse magnetizations of the kink and neutron guide remanent states in small magnetic fields. In order for the polarizer to have a high negative polarization of the output beam, it was proposed to install an additional element between the kink and the neutron guide - $\pi$-neutron spin rotator or a spin-flipper constantly operating in the "*on*" position.

Figure 15 shows a scheme of this polarizer with a kink, a spin-flipper, and a straight polarizing neutron guide. Using this polarizer scheme, polarization and transmission calculations were performed in the *McStas* program for several experimental facilities of the *PIK* reactor. The calculation results presented in [5] demonstrated high output beam characteristics for all considered facilities. In [5], this polarizer was named *TRUNPOSS*. It has been shown that this is a working scheme for the formation of polarized beams in a wide range of neutron wavelengths. However, upon detailed consideration of this version of the polarizer, it was found that for the spin-flipper to work correctly, taking into account the external magnetic fields from the kink and the neutron guide, the distance between the parts of the polarizer must be significant. In this case, the polarizer becomes non-compact, and its total length will be about 1 m. For some experimental facilities, this length of the polarizer becomes unacceptable.

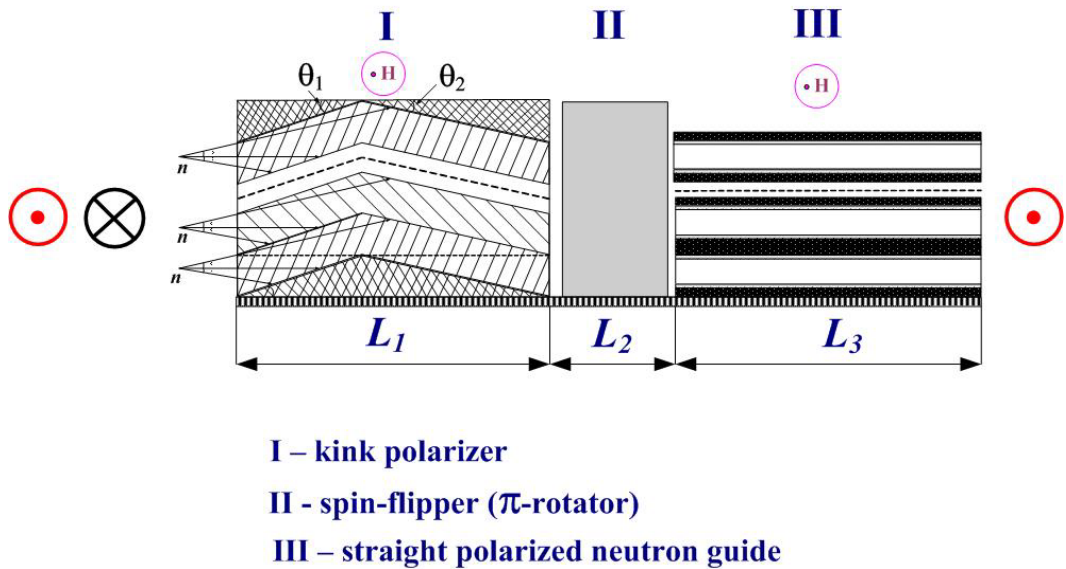

I – kink polarizer
II - spin-flipper ($\pi$-rotator)
III – straight polarized neutron guide

**Fig. 15.** A scheme of a polarizer with a kink, a spin-flipper and a straight polarizing neutron guide. The I and III parts of the polarizer operate in saturating magnetic fields.

**4. Description of the proposed variant of the *SVAROG* transmission polarizer, including the electromagnetic system.**

In this section, we will consider the solution to the problem of creating two remanent states of elements with antiparallel residual magnetization in a transmission polarizer [3, 4]. This was mentioned in section 3.2 of this paper. This problem was solved by using an electromagnet in the polarizer. Figure 16 shows a solenoid with a magnetic yoke made from armko iron with a thickness of 10 mm. The magnetic field in the solenoid is directed along the *z* axis. The length of the solenoid coil along the *x*-axis is 120 mm. In Fig. 17 shows graphs of the spatial distributions of the magnetic field components inside and outside this solenoid. As follows from Fig. 17, inside the solenoid, the magnitude of the magnetic field component directed along the *z* axis is 500 Gs. Outside the solenoid, the value of this component does not exceed -0.9 Gs.

The field components directed along the $x$ and $y$ axes are negligible, both outside and inside the solenoid, i.e. $B_x$ and $B_y \sim 0$. Therefore, two such solenoids with oppositely directed fields can be used sequentially located along the beam along the $x$ axis. Inside these solenoids there will be a kink and a straight polarizing neutron guide. In this case, the neutron beam will pass through the coils of these solenoids. Here, the solenoids can be installed close to each other, since they have no external fields. In this version of the polarizer, there is no need to use a spin-flipper between the kink and the neutron guide.

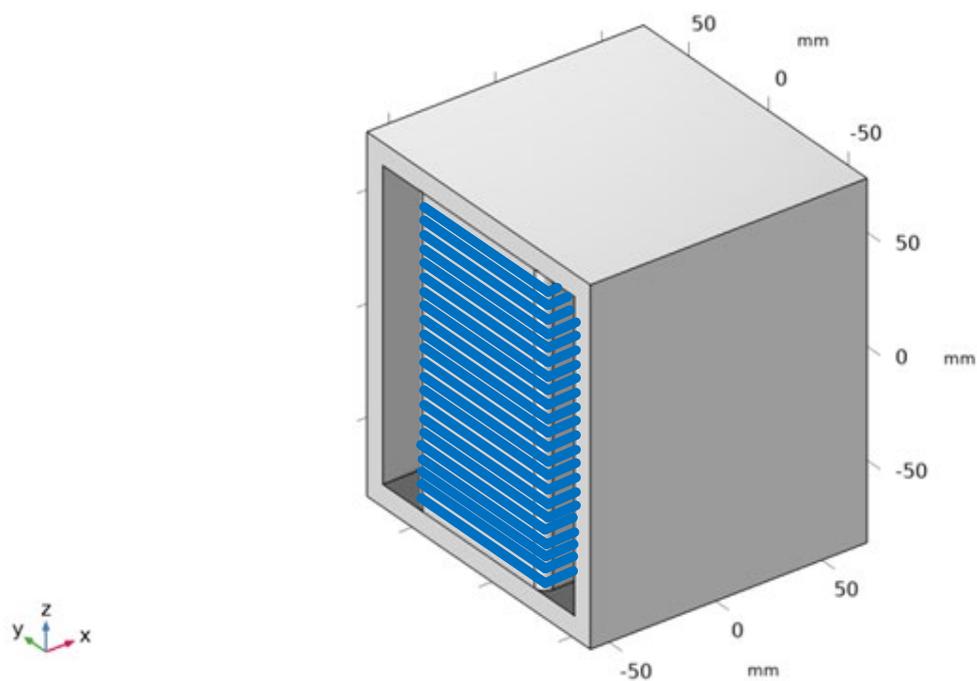

**Fig. 16**. A solenoid with a magnetic yoke made from armko iron.

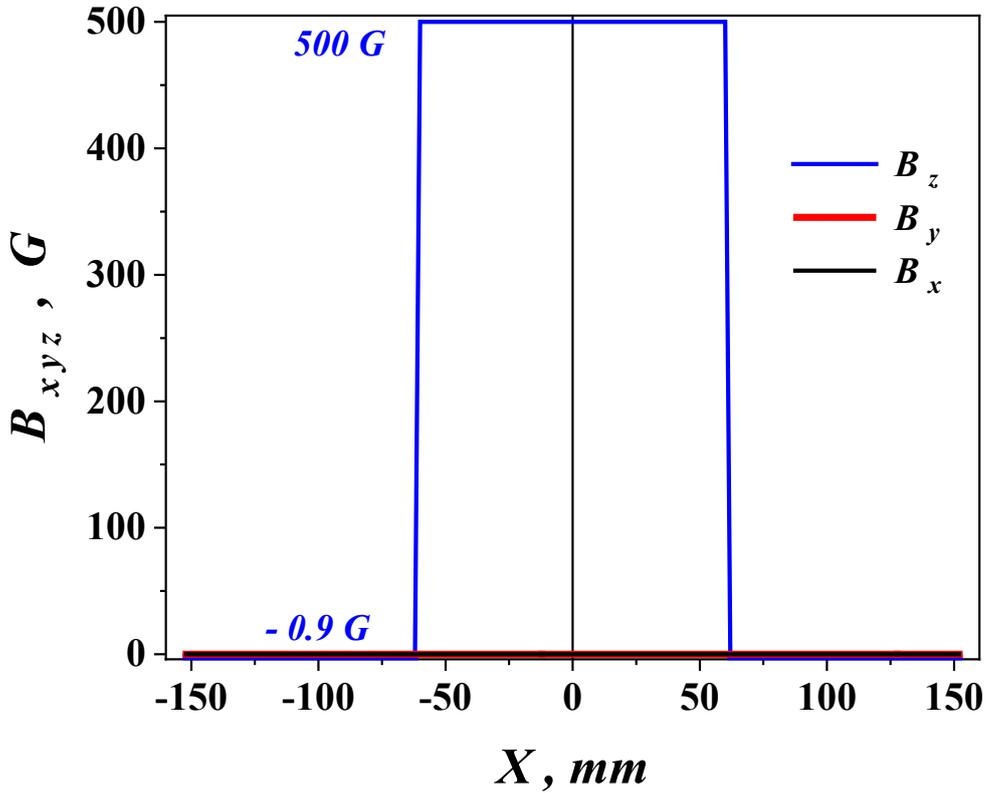

**Fig. 17.** Graphs of spatial distributions of magnetic field components outside and inside the solenoid. Inside and outside the solenoid $B_x$ and $B_y \sim 0$. The value of $B_z$ inside the solenoid is 500 $Gs$, outside the solenoid it does not exceed -0.9 $Gs$.

The optical scheme of the new version of the polarizer, which does not show the electromagnetic system, looks the same as the scheme for the remanent polarizer shown in Fig. 12.

Figure 18 shows top view of the electromagnetic system of the new version of the polarizer. It consists of two coils of solenoids *1* and *2* with lengths $L_1$ and $L_2$, respectively. The coils use aluminum wire, through which a neutron beam passes. Electric currents $I_1$ and $I_2$ in the coils create oppositely directed fields $H_1$ and $H_2$. Their directions are perpendicular to the plane of the drawing. The magnetic yoke of the solenoids is not shown in this figure. As shown in the figure, an unpolarized beam enters the input of the first coil, and only neutrons come out of the second coil, whose spins are parallel to the field vector in this coil. Thus, the beam passed through the polarizer becomes polarized. Figure 19 shows a side view of the electromagnetic system of the new version of the polarizer with common magnetic yokes (*1*) above and below. The figure shows: $H_0$, $H_1$ and $H_2$ are the guiding field, the field in the first coil and in the second coil, respectively; *2* and *3* are the turns of the first and second coils, respectively; *4-1* and *4-2* are the kink wafers; *5* are the of the straight polarizing neutron guide. The compact neutron transmission supermirror polarizer in this version is called **_SVAROG_**.

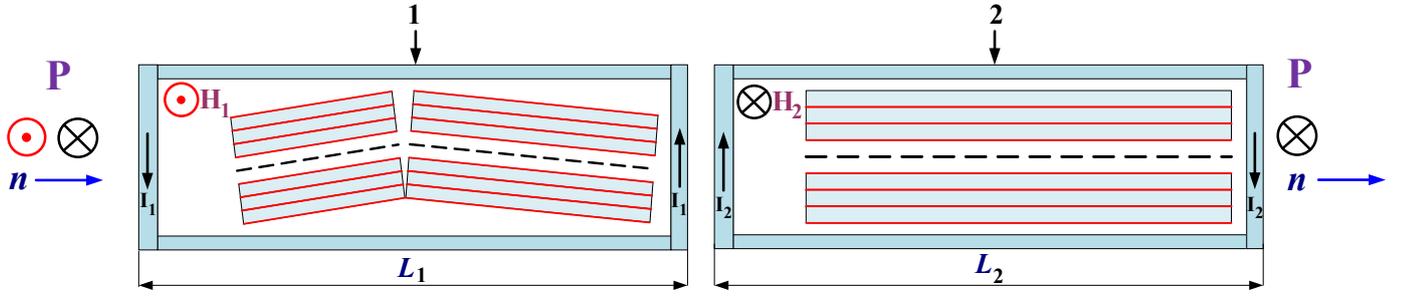

**Fig. 18.** The *SVAROG* polarizer with an electromagnetic system. Top view.

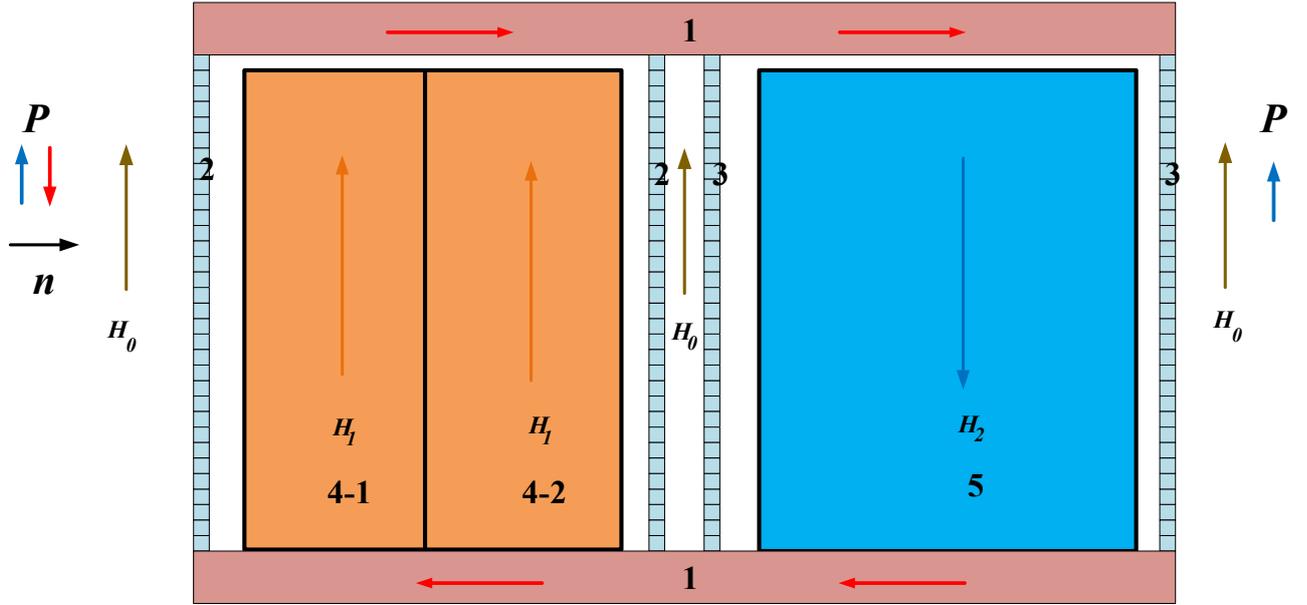

**Fig. 19.** The *SVAROG* polarizer with an electromagnetic system. Side view. $H_0$, $H_1$, and $H_2$ are the guiding field, the field in the first coil and in the second coil, respectively; *1* are the magnetic yokes; *2* and *3* are the turns of the first and second coils, respectively; *4-1* and *4-2* are the kink wafers; *5* are the wafers of the straight polarizing neutron guide.

The process of passing a neutron beam through the coils of both solenoids when electric currents $I_1$ and $I_2$ flow through them was considered. Due to the interaction of neutron spins with the inter-turn magnetic fields of the solenoids, beam depolarization occurs. Therefore, it is necessary to optimize the parameters of the solenoid coils in order to minimize depolarization in order to obtain a highly polarized beam at the exit of the polarizer. Calculations of the distribution of inter-turn magnetic fields depending on the parameters of the aluminum wire and the magnitude of the electric currents flowing through the coils were carried out using the ***COMSOL Multiphysics 6.2*** program [13]. A brief description of this program is given in ***Appendix A***.

Further, taking into account the pattern of the distribution of inter-turn fields, calculations of the depolarization of the beam with a divergence of ± 0.25 degrees in the horizontal plane during its passage through the coils of the solenoids using the *McStas* program were carried out. Figure 20 shows the spectral dependences of the depolarization of a neutron beam passing through the polarizer solenoids for two types of wire with an electrical insulation thickness of 0.1 mm: a round wire with a diameter of 2 mm (blue curve) and a wire with a square section of 2 mm x 2 mm (red curve). The distance between the coils of the solenoids is 1 mm. Oppositely directed fields with a field strength of 500 *Gs* were created in the solenoids. As follows from the graph, the depolarization of the beam strongly depends on the shape of the wire section. At the same time, a square wire section is noticeably better than a round one. The results of the process of minimizing the depolarization of a neutron beam during its passage through two solenoids, depending on their parameters (wire cross-section and shape, electrical insulation thickness, distance between coils) are presented in *Appendix A*. The *Appendix A* contains the results of calculations of the spectral dependences of depolarization obtained by optimizing the parameters of the solenoids.

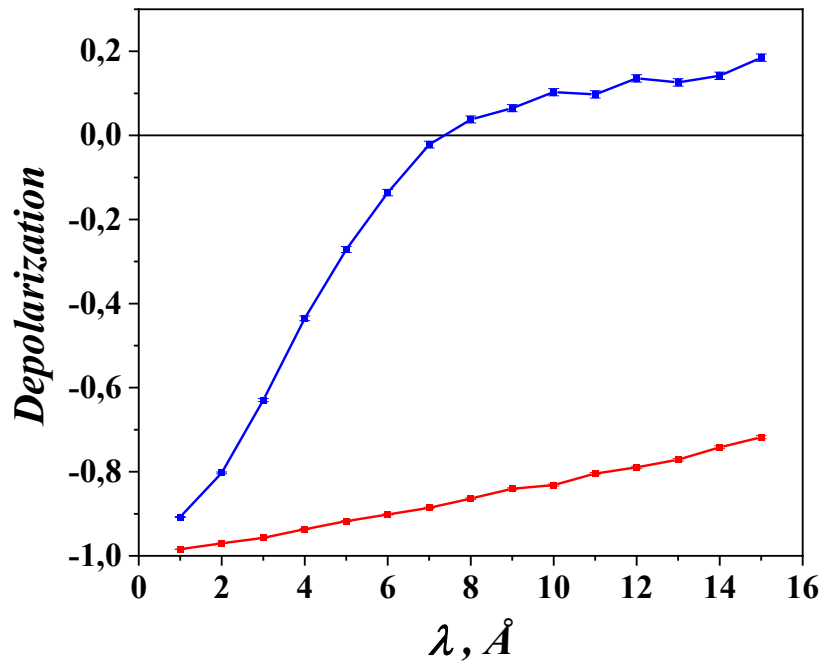

**Fig. 20.** Spectral dependences of the depolarization of a neutron beam passing through polarizer solenoids for two types of wire with an electrical insulation thickness of 0.1 mm: a round wire with a diameter of 2 mm (blue curve) and a square wire with a cross section of 2 mm x 2 mm (red curve). The distance between the coils of the solenoids is 1 mm. Oppositely directed fields of 500 *Gs* were created in the solenoids.

Figures 21a, b show the distribution of the components of the magnetic fields along the beam axis, obtained for optimized parameters of the electromagnetic system: a square wire with a cross-section of 2 mm x 2 mm, a wire electrical insulation thickness of 0.02 mm, and a distance between coils of 6 mm. Fields: in the 1st coil +85 *Gs*, in the 2nd coil -114 *Gs*, the guiding field between the coils, at the input and output of the 2nd coil +15 *Gs*. The guiding fields appear due to unequal fields in the coils.

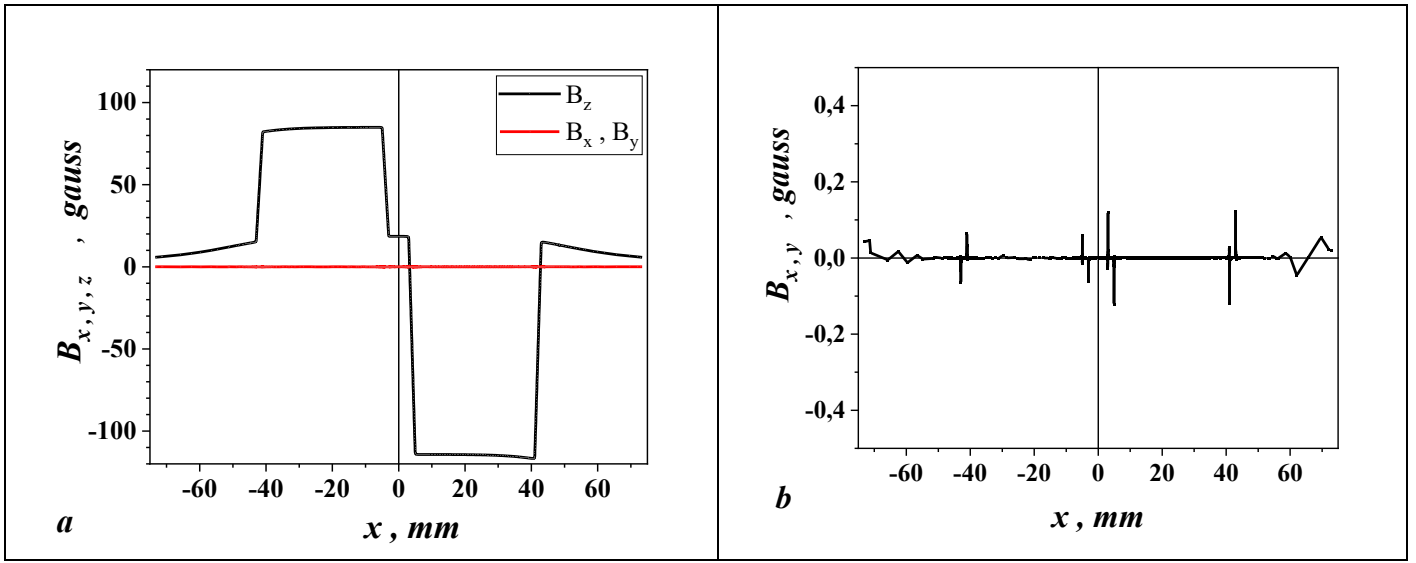

**Fig. 21a, b.** Distribution of magnetic field components obtained for optimized electromagnetic system parameters.

As can be seen from Fig. 21, there is only one component of the $B_z$ field in the solenoids (the values of the other two components do not exceed 0.1 *Gs*): in the first coil, its value is +85 *Gs*, in the second -114 *Gs*. In addition, guiding fields (+15 *Gs*) are created between the coils, at the input and exit of the 2nd coil, which ensures that the beam does not depolarize in these areas of the polarizer. The beam travels the distance between the coils non-adiabatically due to a sharp change in the field. It is assumed that at the beginning of work with the polarizer, maximum fields (+500 *Gs* and -500 *Gs*) are created for a short time, and then their values are reduced to operating values: +85 *Gs* and -114 *Gs*. At the same time, as discussed above, due to the remanence property of supermirrors, their polarizing efficiency practically does not decrease. Since the values of the operating fields are significant (+85 *Gs* and -114 *Gs*), therefore, the influence of external scattered magnetic fields on the remanent states of polarizing supermirrors will not be so significant.

Figure 22 shows the final results of calculating the spectral depolarization of a neutron beam passing through the coils of two solenoids. As follows from Fig. 22, there is practically no depolarization of the beam when passing through both solenoids with optimized parameters.

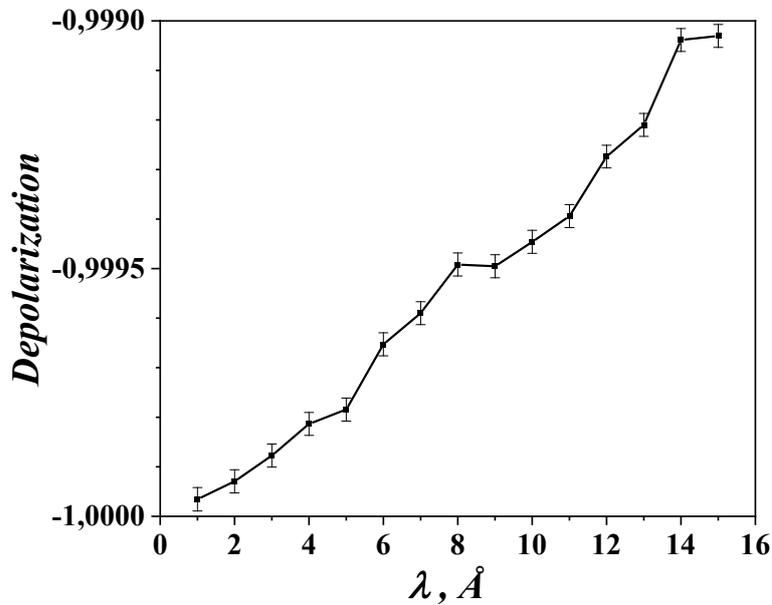

**Fig. 22.** The final results of calculating the spectral depolarization of a neutron beam passing through the coils of two solenoids with optimized parameters.

The papers [14-17] present the results of successful work with solenoids, through coils of which a neutron beam passes, the depolarization of which is minimal. The turns of these solenoids were rectangular aluminum plates. As a variant, a solenoid is considered, the turns of which are aluminum plates of rectangular cross-section. Figures 23 and 24 show the top and front views of such a turn. Table 2 shows the parameters of a solenoid coil consisting of rectangular aluminum plates shown in Figures 23 and 24. As follows from Table 2, with nominal values of 114 *Gs* and 85 *Gs* fields, the heat dissipation in the coils is very low and is equal to 29 watts and 16 watts, respectively. With such heat generation, no special cooling is required. The heat generation in the coils can be neglected when the maximum field of 500 *Gs* is switched on for a short time.

Table 2. Parameters of a solenoid coil consisting of aluminum turns - plates of rectangular cross-section:

Dimensions of turns (plates) (LxHxW): 130 mm x 140 mm x 80 mm

Number of turns (plates): 70

The turn is a rectangular plate: 2 mm x 2 mm long 2 x 30 mm; 2 mm x 6 mm long 360 mm

Plate Material: Aluminum

Electrical insulation: dacron film 20 microns thick

Total resistance of the solenoid coil: 0.0882 ohms

| Field , *Gs* | I , *Ampere* | U , *volt* | P , *watts* |
|---|---|---|---|
| **500** | 79,5 | 7,0 | 557 |
| **114** | 18,13 | 1,6 | 29 |
| **85** | 13,5 | 1,2 | 16 |

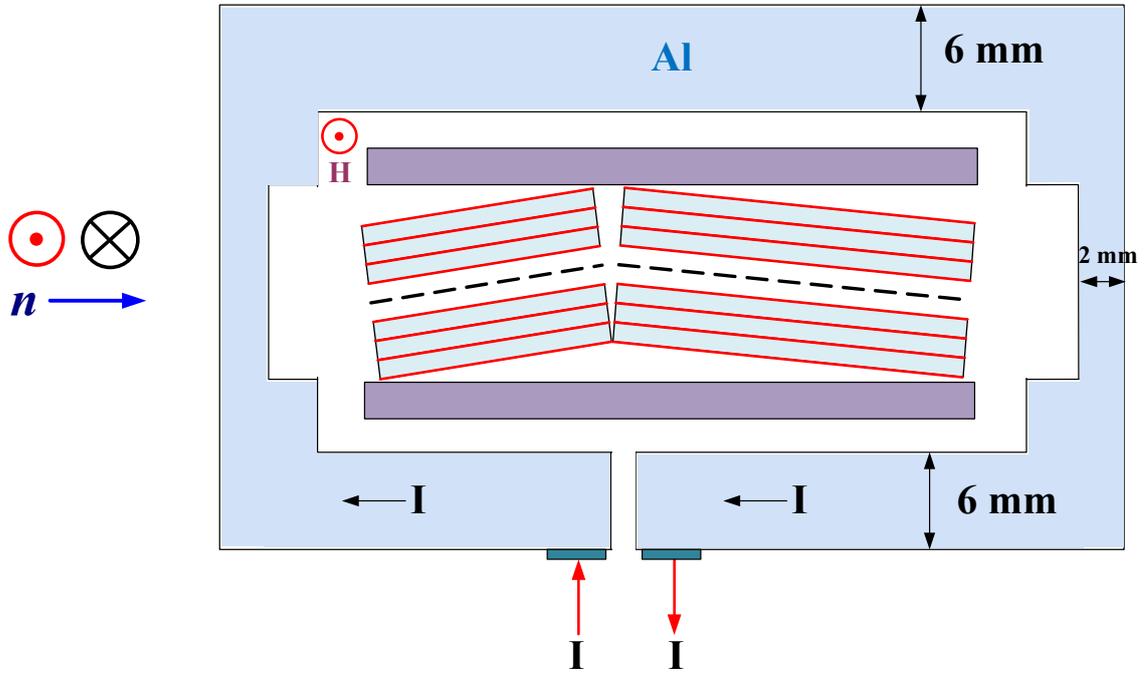

**Fig. 23.** Top view of an aluminum plate representing a turn of a solenoid coil.

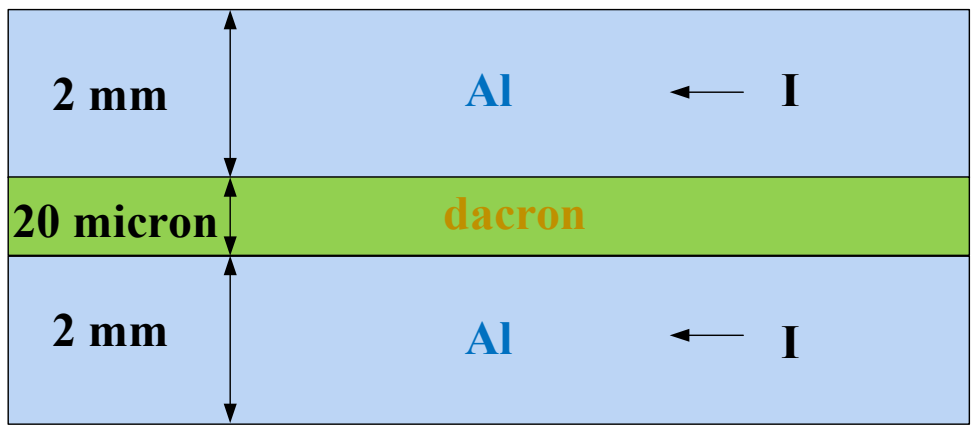

**Fig. 24.** Front view of an aluminum plate representing a turn of a solenoid coil.

## 5. Description of calculations of the intensity of both spin components of the beam passed through the *SVAROG* polarizer.

The intensities of the (+) and (-) spin components of the beam passed through the polarizer were calculated using the **Particle Raytracing** program [18]. A brief description of this program is given in *Appendix B*. Let us consider in detail the process of passing through the polarizer of both spin components of a neutron beam with a wavelength of 5 Å. Polarizer parameters (see Fig. 12): supermirror coating with parameter $m = 2.5$, silicon wafer thickness $d = 0.3$ mm; wafer lengths: $L_1 = 20$ mm, $L_2 = 30$ mm, $L_3 = 120$ mm; angles: $\theta_1 = 0.859$ degrees, $\theta_2 = 0.573$ degrees. Input beam parameters: angular divergence $\pm 0.45$ degrees, width 30 mm.

Figures 25-28 show the calculated angular intensity dependences for both spin components of the beam at the entrance to the polarizer (Fig. 25), after a stack of wafers of length $L_1$ (Fig. 26), after a stack of wafers of length $L_2$ (Fig. 27), at the exit of the polarizer or after a stack of wafers of length $L_3$ (Fig. 28), respectively.

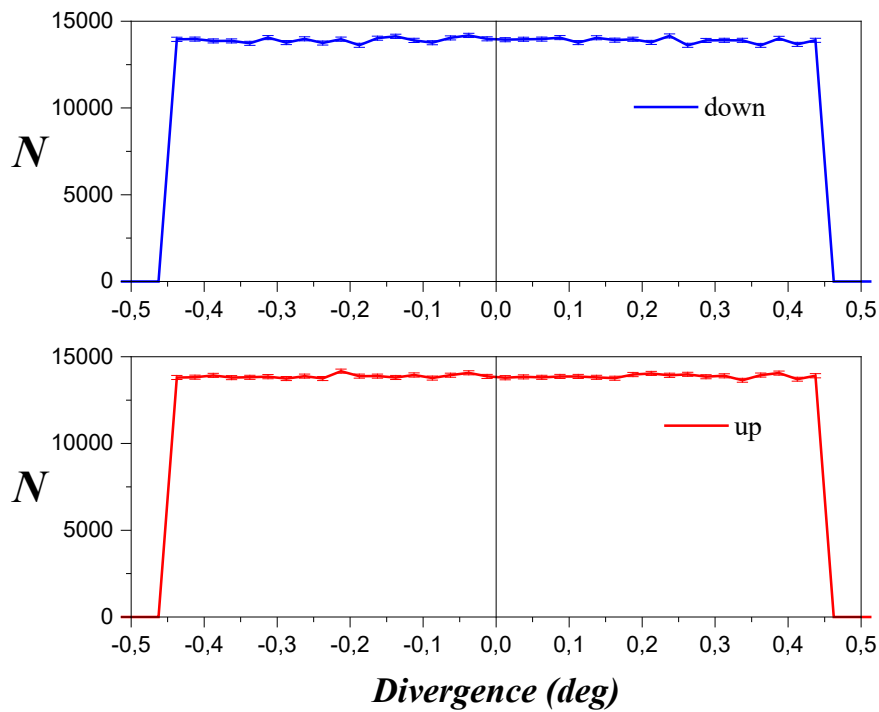

**Fig. 25.** Angular intensity dependences for both spin components of the beam at the entrance to the polarizer.

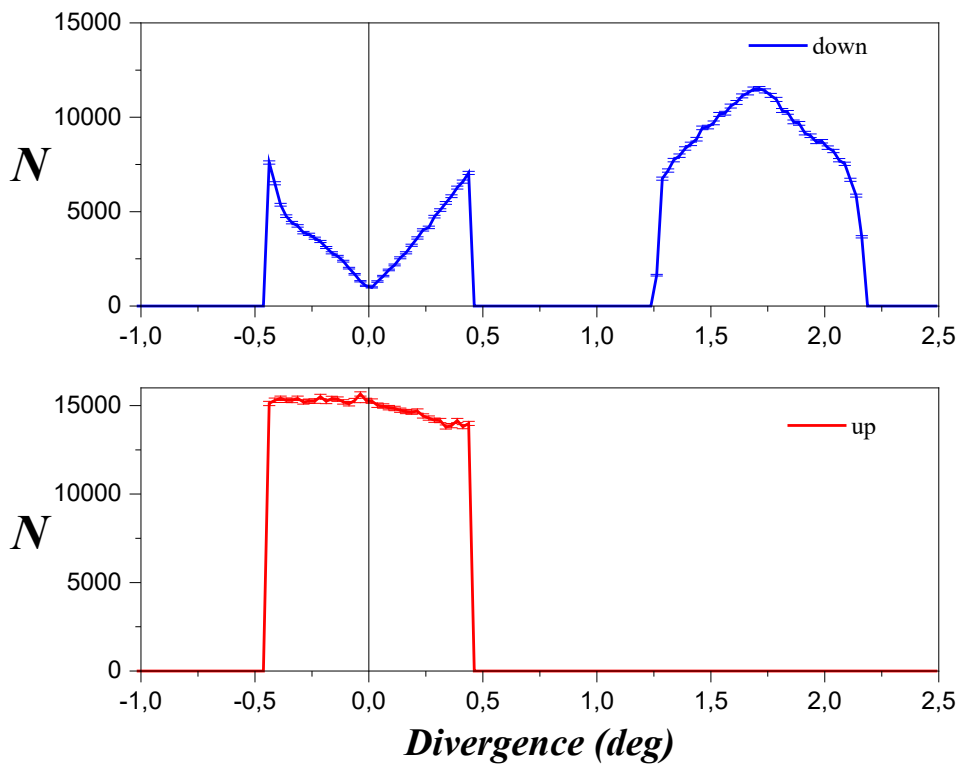

**Fig. 26.** Angular intensity dependences for both spin components of the beam after a stack of wafers of length $L_1$.

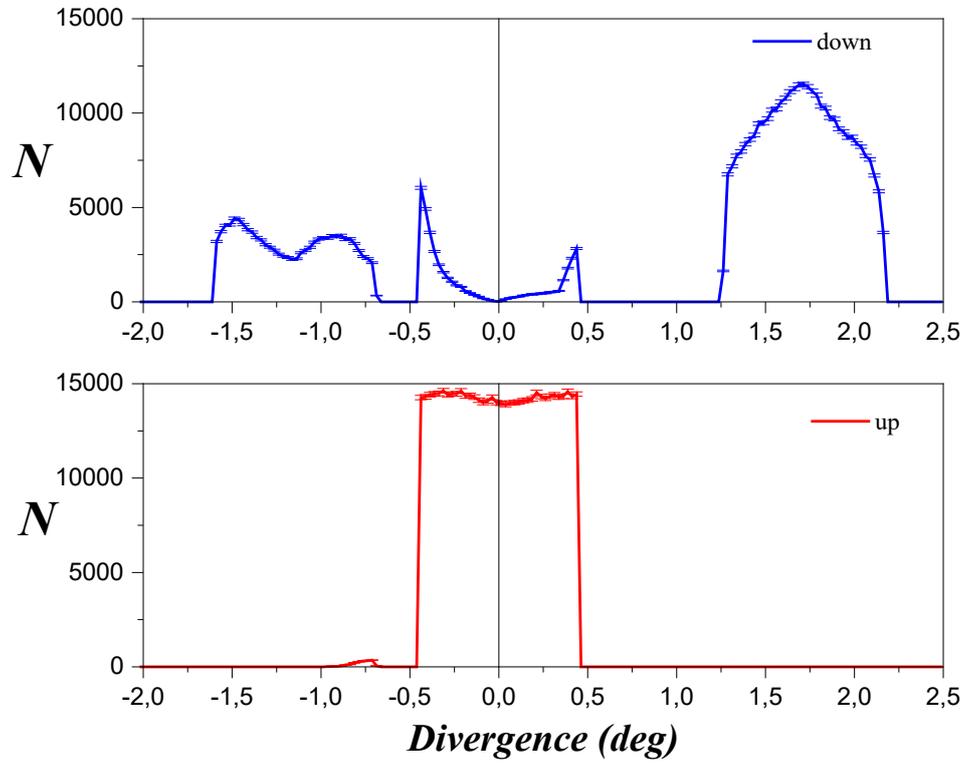

**Fig. 27.** Angular intensity dependences for both spin components of the beam after a stack of wafers of length $L_2$.

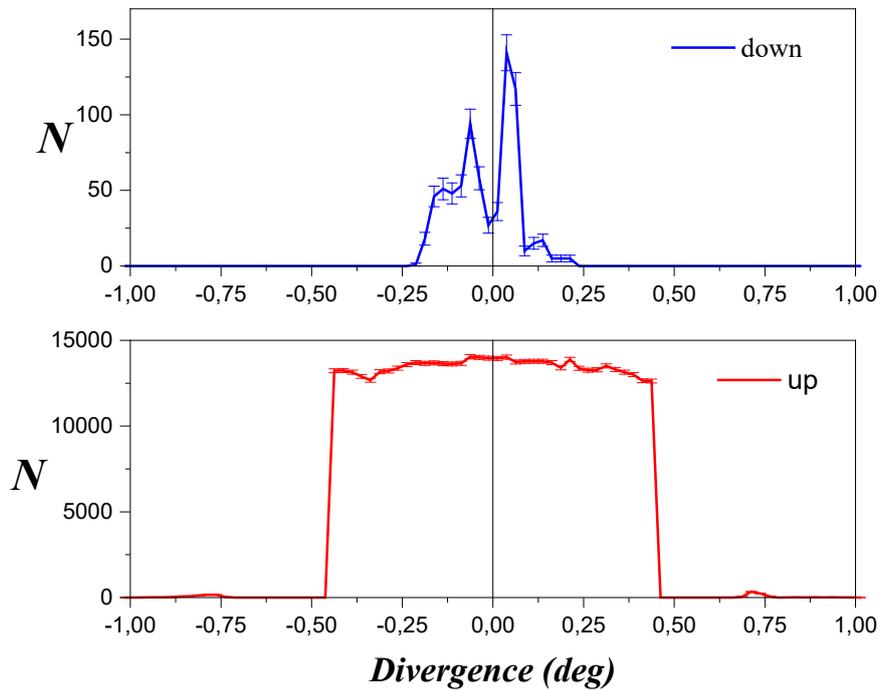

**Fig. 28.** Angular intensity dependences for both spin components of the beam after a stack of wafers with length $L_3$ at the exit of the polarizer.

Three more polarizers with $L_3=120$ mm and with the corresponding sets of parameters $L_1$ and $L_2$ were also considered. Graphs similar to those shown in Figs. 25-28 were constructed for each polarizer and the integral values of the polarization $P$ and the transmission coefficient (-) of the spin component of the $T^-$ beam with a wavelength of 5 Å were obtained from them. Table 3 shows, for four sets of parameters $L_1$ and $L_2$, the integral values of the polarization $P$ and the transmission coefficient (-) of the spin component of the $T^-$ beam with a wavelength of 5 Å passed through a polarizer with $L_3 = 120$ mm.

The value of $P$ was calculated using the formula: $P = (I^+ - I^-)/(I^+ + I^-)$, where the values of $I^+$ and $I^-$ were found by integrating over all angles at the exit of the polarizer (Fig. 28). Values of $T^-$ were calculated according to the formula: $T^- = I^- / I_0^-$, where the value of $I^-$ was found by integration at all angles at the exit of the polarizer (Fig. 28 for the (-) component), and the value of $I_0^-$ was found by integration at all angles at the input of the polarizer (Fig. 25 for the (-) component). As follows from Table 3, non-symmetrical variant No. 1 looks preferable, because the transmission coefficient for it is higher, the length of the plates is shorter, therefore, the absorption of the beam in silicon is lower, while the polarization is slightly less than the similar values of polarization for symmetrical variants.

**Table 3.** Integral values of polarization $P$ and transmission $T^-$ for (-) spin component of the beam with a wavelength of 5 Å through a polarizer with $L_3 = 120$ mm.

| № | Kink | | $P$ | $T^-$ (excluding absorption) |
|---|---|---|---|---|
| 1 | Not Symmetrical | $L_1 = 20$ mm, $L_2 = 30$ mm | 99.69 | 97.35 |
| 2 | Symmetrical | $L_1 = 30$ mm, $L_2 = 30$ mm | 99.91 | 96.99 |
| 3 | Symmetrical | $L_1 = 35$ mm, $L_2 = 35$ mm | 99.92 | 96.96 |
| 4 | Not Symmetrical | $L_1 = 41$ mm, $L_2 = 64$ mm | 97.00 | 97.01 |

Thus, by integrating over the angle, it is possible to obtain the polarization value for each wavelength and the transmission coefficient for each spin component of the beam for this wavelength. Therefore, for a given wavelength range, the spectral dependences P and T - can be obtained. Such calculations were performed for the wavelength range $\lambda = 2.4 - 15$ Å. The neutron physics facilities of the *PIK* reactor's instrument base will operate in this *range: IN2, DEDM, TENZOR, SESANS, SEM. Calculations of the spectral dependences of the polarizer P and $T^-$* were performed for 11 combinations of the kink lengths $L_1$ and $L_2$ at a constant length of the neutron guide $L_3$. The results of calculations of spectral polarization $P$ are shown in Fig. 29. Polarizer parameters: supermirror coating $m = 2.0$, silicon wafer thickness $d = 0.3$ mm, $L_3 = 120$ mm. Input beam: divergence ± 0.45 degrees, width 30 mm.

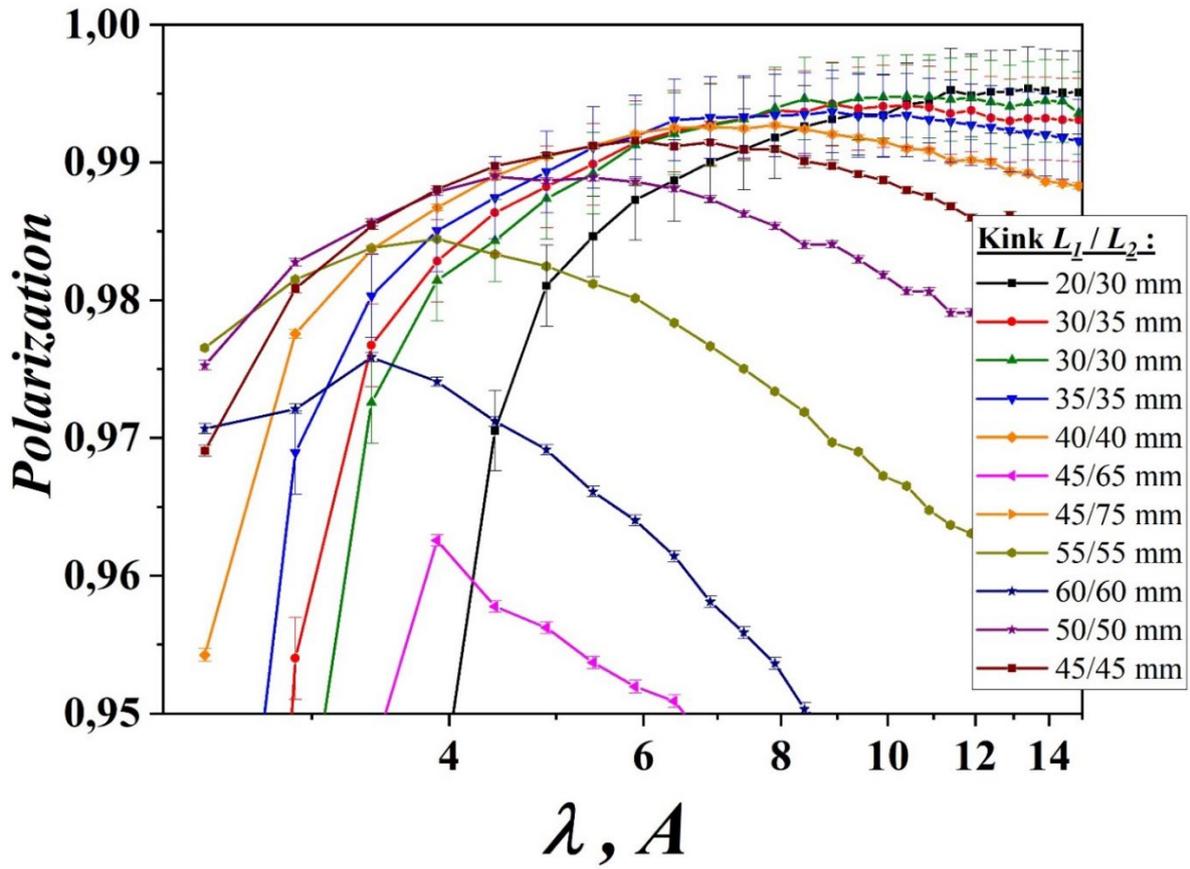

**Fig. 29.** Calculated spectral dependences of the polarization $P$ for combinations of the kink lengths $L_1$ and $L_2$ of the polarizer at a constant value $L_3$.

From the *11* combinations of $L_1$ and $L_2$ values, *4* $L_1/L_2$ combinations were selected: 40/40 mm, 40/45 mm, 45/45 mm and 45/50 mm. For each of these combinations with the same parameters of the supermirror and the beam, the spectral dependences of the polarization $P$ were calculated for four $L_3$ values: 80, 100, 120 and 140 mm. These dependencies are shown in Figures 30-33.

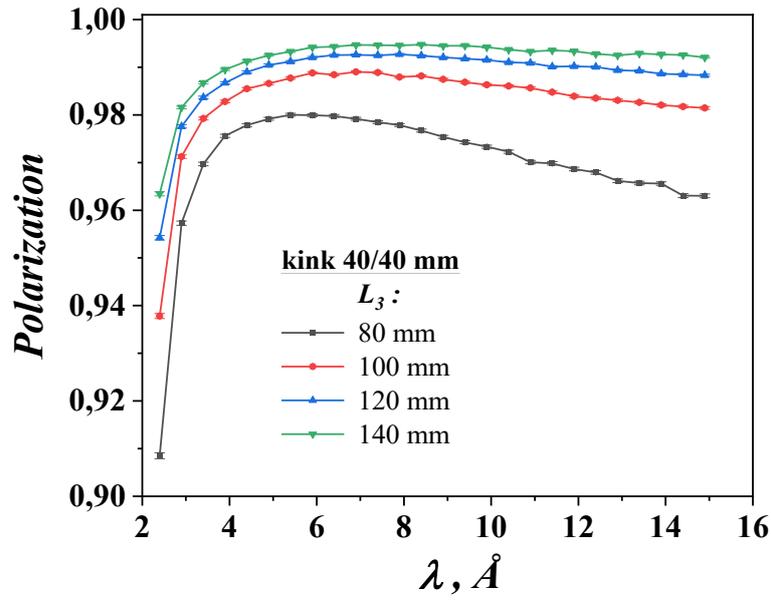

**Fig. 30.** Spectral dependences of the polarization $P$ for kink lengths $L_1 = 40$ mm and $L_2 = 40$ mm for four lengths $L_3 = 80, 100, 120$ and $140$ mm.

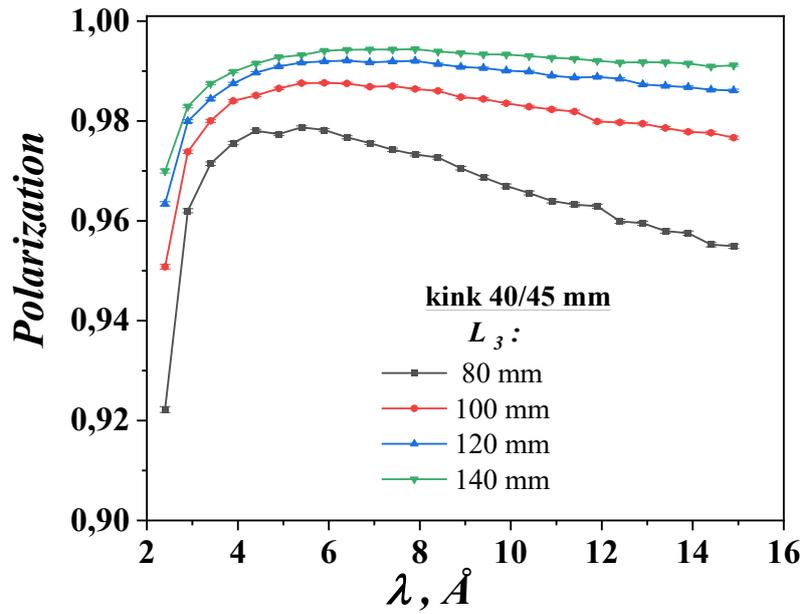

**Fig. 31.** Spectral dependences of the polarization $P$ for kink lengths $L_1 = 40$ mm and $L_2 = 45$ mm for four lengths $L_3 = 80, 100, 120$ and $140$ mm.

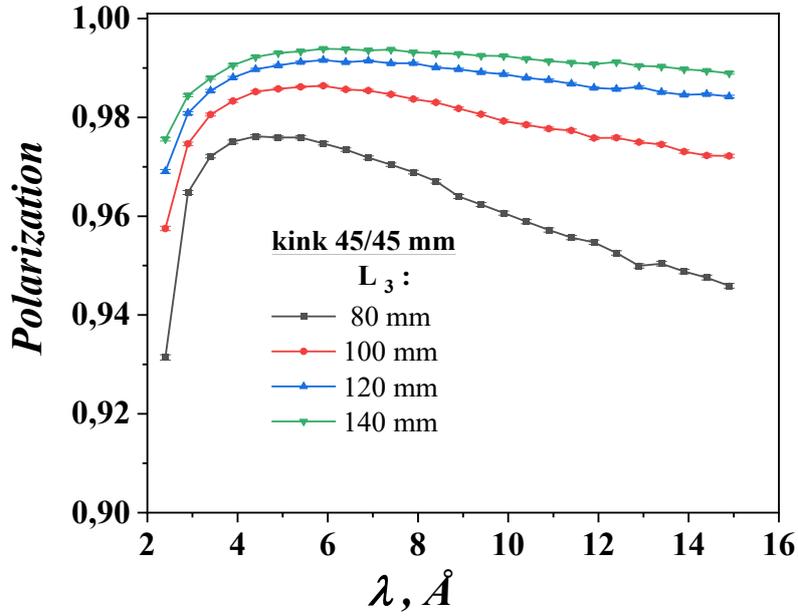

**Fig. 32.** Spectral dependences of the polarization $P$ for kink lengths $L_1 = 45$ mm and $L_2 = 45$ mm for four lengths $L_3 = 80, 100, 120$ and $140$ mm.

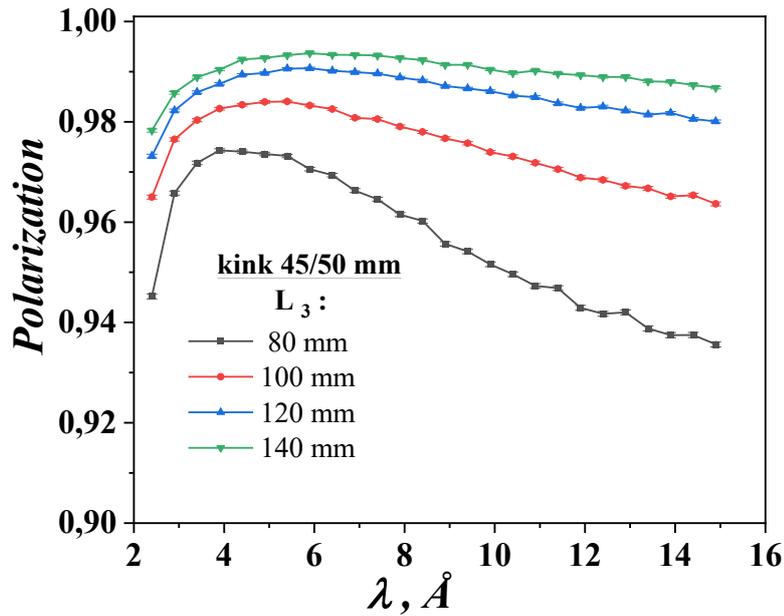

**Fig. 33.** Spectral dependences of the polarization $P$ for kink lengths $L_1 = 45$ mm and $L_2 = 50$ mm for four lengths $L_3 = 80, 100, 120$ and $140$ mm.

Based on the calculations performed, the optimal polarizer variant was selected with kink lengths: $L_1 / L_2 = 40$ mm / $45$ mm and neutron guide length: $L_3 = 100$ mm. For this variant, the total length of the polarizer is minimal and, taking into account the length of the electromagnetic system, is 240 mm, therefore, beam losses in silicon are minimal, and the polarization of the output beam from the polarizer exceeds 0.95 over the entire wavelength range $\lambda = 2.4 – 15$ Å.

Figure 34 shows the spectral dependences of $P$ and $T^-$ for a polarizer with optimal parameters for the wavelength range $\lambda = 2.4 – 15$ Å for the lengths of the kink plates $L_1 = 40$ mm, $L_2 = 45$ mm and the length of the neutron guide $L_3 = 100$ mm with a divergence of the input beam ± 0.45 degrees. When calculating the $T^-$ absorption of neutrons in silicon, it was not taken into account.

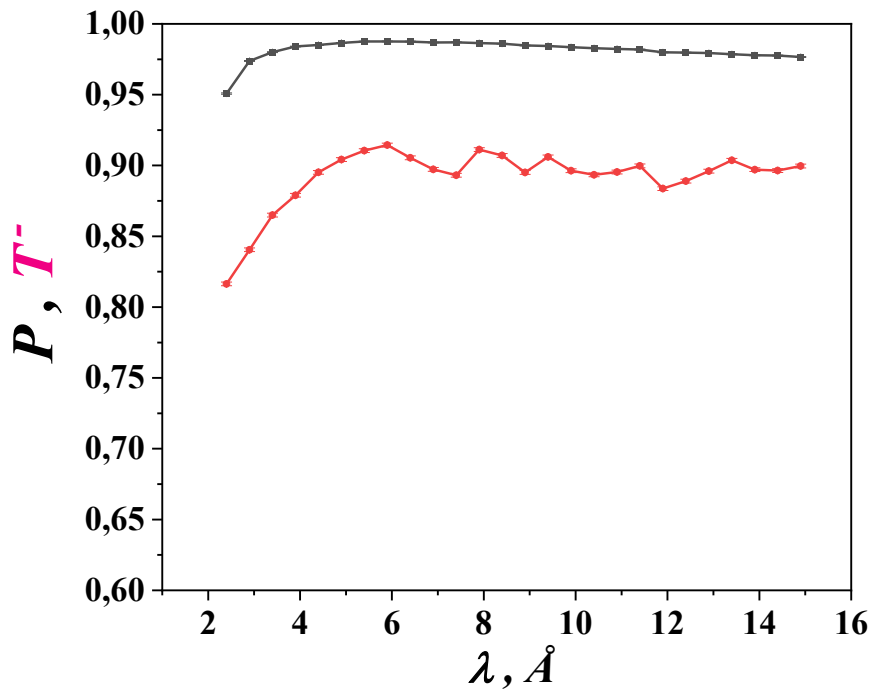

**Fig. 34.** Spectral dependences of $P$ and $T^-$ for a polarizer with optimal parameters for the wavelength range $\lambda = 2.4 – 15$ Å for the kink lengths $L_1 = 40$ mm, $L_2 = 45$ mm and the length of the neutron guide $L_3 = 100$ mm with a divergence of the input beam ± 0.45 degrees.

Figure 35 shows the calculated spectral dependences of $P$ and $T^-$ for a polarizer with optimal parameters for lengths $L_1 = 40$ mm, $L_2 = 45$ mm, and $L_3 = 100$ mm for three input beam angular divergence values: ± 0.3, ± 0.45, and ± 0.6 degrees. As follows from the graph, the polarization practically does not change with increasing input angular divergence. The transmission coefficient at small wavelengths decreases with increasing input angular divergence. Although at the same time, the absolute intensity of the beam does not decrease for these wavelengths.

Figure 36 shows the calculated spectral dependences of $P$ and $T^-$ for a polarizer with optimal parameters for the lengths $L_1 = 40$ mm, $L_2 = 45$ mm and $L_3 = 100$ mm for two values of the parameter of the supermirror coating: $m = 2.0$ and $m = 2.5$ and the divergence of the input beam ± 0.45 degrees. As follows from the graph, the polarization is practically independent of the parameter $m$. The average spectral transmission coefficient for the parameter $m = 2.5$ is about 5% higher than the same coefficient for the parameter $m = 2.0$.

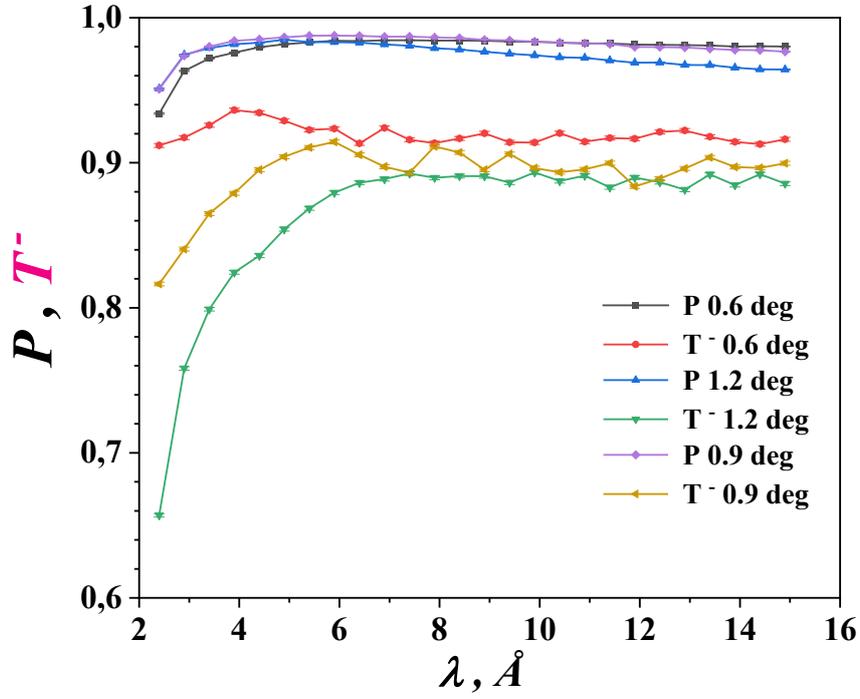

**Fig. 35.** Spectral dependences of $P$ and $T^-$ for a polarizer with optimal parameters for lengths $L_1$ = 40 mm, $L_2$ = 45 mm and $L_3$ = 100 mm for three values of total angular divergence of the input beam: 0.6, 0.9 and 1.2 degrees.

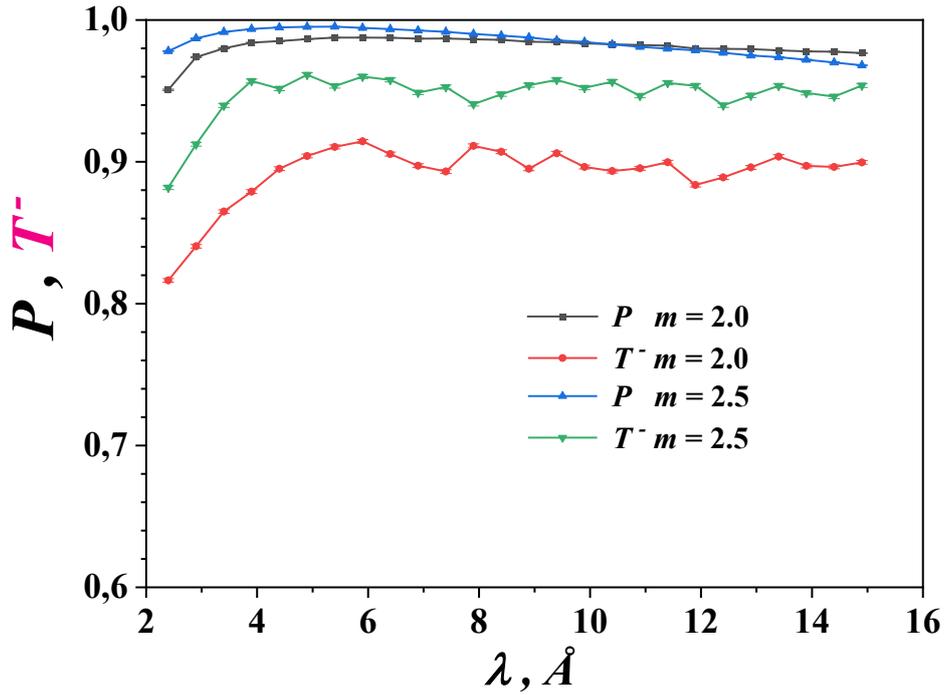

**Fig. 36.** Spectral dependences of $P$ and $T^-$ for a polarizer with optimal parameters for the lengths $L_1$ = 40 mm, $L_2$ = 45 mm and $L_3$ = 100 mm for two values of the parameter of the supermirror coating: $m$ = 2.0 and $m$ = 2.5 and the total angular divergence of the input beam 0.9 degrees.

Figures 37 and 38 show the calculated angular dependences of the intensities of both spin components of the beam at the input and exit of the polarizer with optimal parameters for a wavelength of 2.4 Å, respectively. The integral polarization of the beam at the output of the polarizer is 0.951.

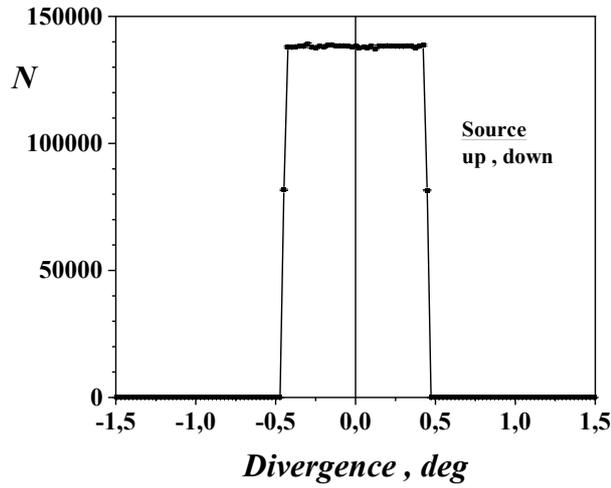

**Fig. 37.** Calculated input angular dependences of the intensities of both spin components of the beam for a polarizer with optimal parameters for a wavelength of 2.4 Å.

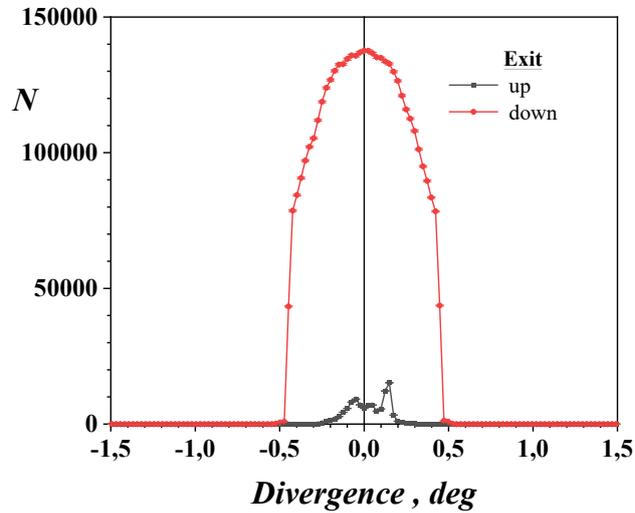

**Fig. 38.** Calculated exit angular dependences of the intensities of both spin components of the beam for a polarizer with optimal parameters for a wavelength of 2.4 Å.

For the wavelength range $\lambda = 0.9 – 2.4$ Å, the optimal polarizer parameters were obtained in the same way as for the wavelength range $\lambda = 2.4 - 15$ Å. The optimal parameters of polarizer for the wavelength range $\lambda = 0.9 – 2.4$ Å: the *CoFe/TiZr* parameter of the supermirror coating $m = 2.5$, the thickness of the silicon wafers $d = 0.15$ mm, the wafer lengths $L_1 = 44$ mm, $L_2 = 60$ mm, $L_3 = 100$ mm. The total length of the polarizer, taking into account the length of the magnetic system, is 260 mm. Input beam: angular divergence ± 0.20 degrees, width 30 mm. The *IN3* and *D3* neutron physics facilities of the *PIK* reactor will operate in this wavelength range. In Fig. 39 shows the calculated spectral dependences of the polarization $P$ and the transmission of the (-) spin component of the $T^{\,-}$ beam at the exit of the polarizer with optimal parameters for the wavelength range $0.9 – 2.4$ Å. The beam losses in silicon were not taken into account here. The polarization of the output beam from the polarizer exceeds 0.95 over the entire wavelength range $\lambda = 0.9 – 2.4$ Å.

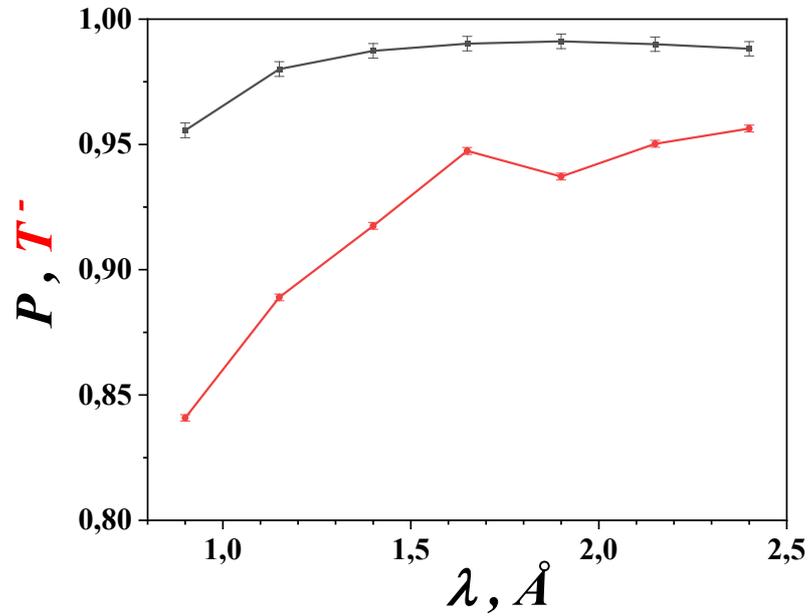

**Fig. 39.** Spectral dependences of polarization $P$ and transmission $T^-$ of the (-) spin component of the beam for the wavelength range 0.9 – 2.4 Å at the exit of the polarizer with optimal parameters for lengths $L_1 = 44$ mm, $L_2 = 60$ mm and $L_3 = 100$ mm at an angular divergence of the input beam ± 0.2 degrees.

Figures 40 and 41 for a wavelength of 1.5 Å show the calculated angular dependences of the intensities of both spin components of the beam at the entrance and exit of the polarizer with optimal parameters. The integral polarization of the beam at the exit of the polarizer is high and equal to 0.989.

*Appendix B* presents calculations of the polarization and transmission of the beam at the exit of the polarizer depending on the magnitude of deviations from the correct geometry of the kink (shifts and offsets, angular dispersion), as well as the dependence of the total neutron beam scattering cross section in silicon and aluminum.

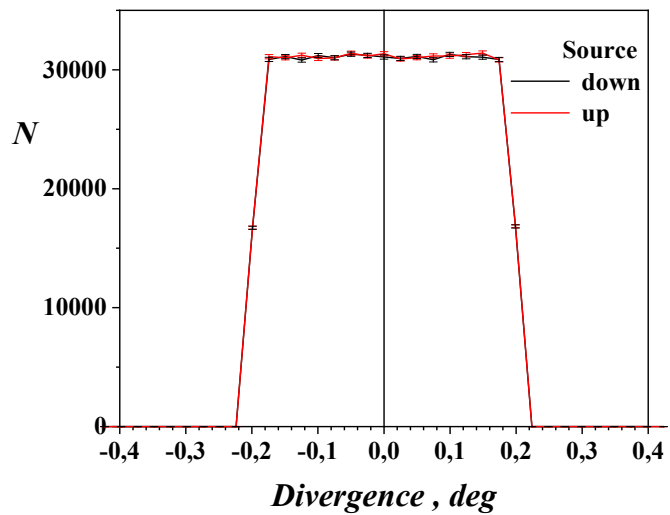

**Fig. 40.** Calculated input angular dependences of the intensities of both spin components of the beam for a wavelength of 1.5 Å for a polarizer with optimal parameters.

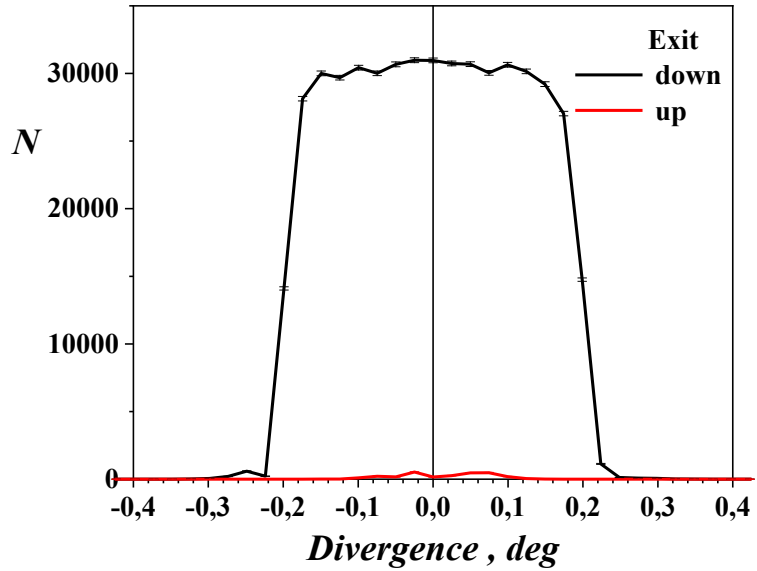

**Fig. 41.** Calculated exit angular dependences of the intensities of both spin components of the beam for a wavelength of 1.5 Å for a polarizer with optimal parameters.

# 6. Calculations of the spectral dependences of polarization and beam transmission for the *SVAROG* polarizer with air channels, taking into account the absorption and scattering of the beam in silicon and aluminum.

In [5], the scheme of the *SVAROG* transmission polarizer with air channels was considered. Figure 42 shows a top view of this scheme along with schemes of wafers of the kink and the neutron guide. The difference between this polarizer scheme and the one discussed above in Fig. 12 is that in both parts of the polarizer, every second wafer is replaced by an air gap - a channel with a width equal to the thickness of the wafer. In addition, a supermirror coating is sputtered twice on each side of the silicon wafer of the $L_3$ *CoFe/TiZr* neutron guide and an absorbing layer is sputtered between these coatings. The supermirror coating of the wafers with lengths $L_1$ and $L_2$ is the same as in Fig. 12.

This version of the polarizer is cheaper, because the number of wafers for the same cross-beam section requires 2 times less. In addition, its luminosity is noticeably higher than that of a polarizer without air gaps, since the average path length of neutrons in silicon in this polarizer will be two times less. For the wavelength range of 2.4 – 15 Å, the optimal parameters of this polarizer were the same as for the polarizer shown in Fig. 12: the *CoFe/TiZr* parameter of the supermirror coating is $m = 2.0$, the thickness of the silicon wafers and air channels is $d = 0.30$ mm, the wafer lengths are $L_1 = 40$ mm, $L_2 = 45$ mm, $L_3 = 100$ mm. Input beam: angular divergence ± 0.45 degrees, beam width 30 mm.

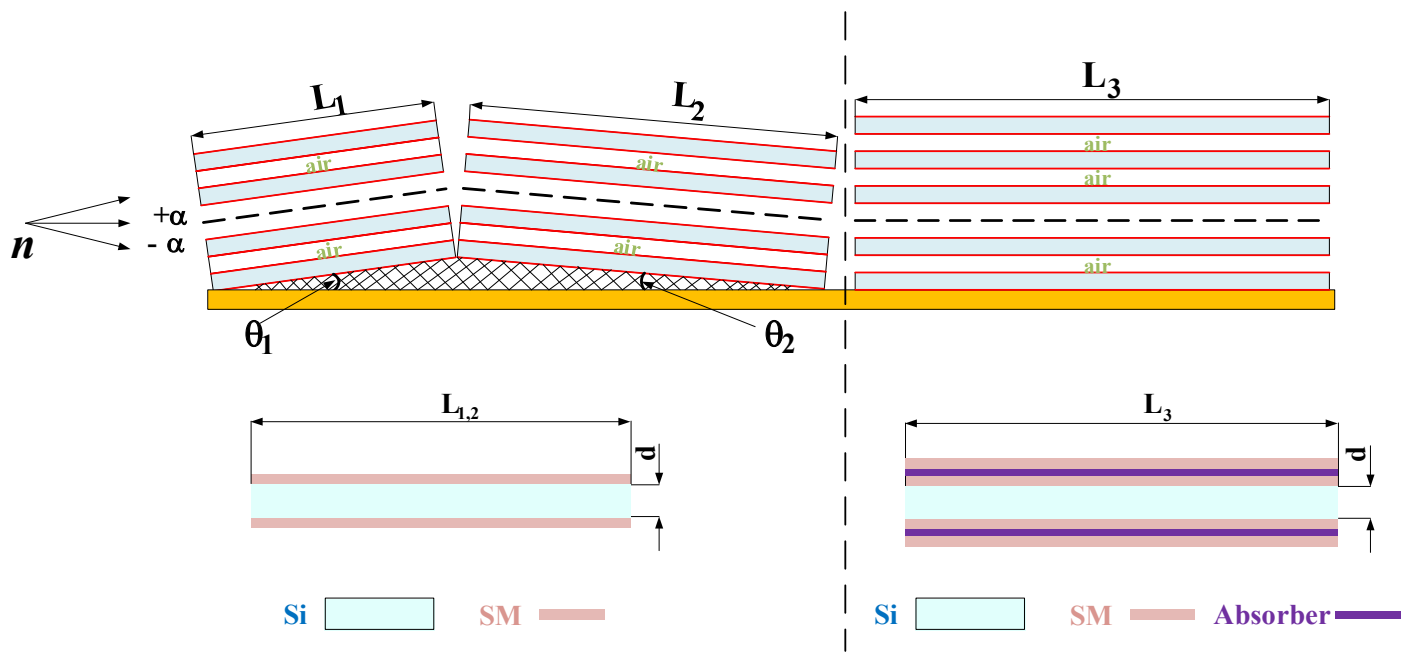

**Fig. 42.** The scheme of the *SVAROG* polarizer, in which silicon wafers alternate with air gaps-channels with a width equal to the thickness of the wafers. Top view. The figure shows also the wafers of the kink and the neutron guide. The electromagnetic systems of kink and neutron guide are not shown in the figure.

Figure 43 shows the calculated spectral dependences of the polarization $P$ and the transmission coefficient $T^-$ for (-) spin component of the beam at the exit of the polarizer with optimal parameters for the wavelength range 2.4 – 15 Å. The spectral transmission of $T^-$ is represented by three curves: a curve without absorption and two curves with absorption: a polarizer without air channels and a polarizer with air channels. The absorption and scattering of neutrons in silicon was taken into account in accordance with [19]. The graph from this work is given in *Appendix B*. As follows from Fig. 43, the polarization of the exit beam is high and exceeds 0.95. The transmission coefficient $T^-$ is also high for a polarizer with air channels and its average value for the wavelength range $\lambda$ = 2.4 - 6 Å is 0.69, which exceeds the same value for a polarizer without air channels by 28%. Here, the absorption of the beam in aluminum was not taken into account.

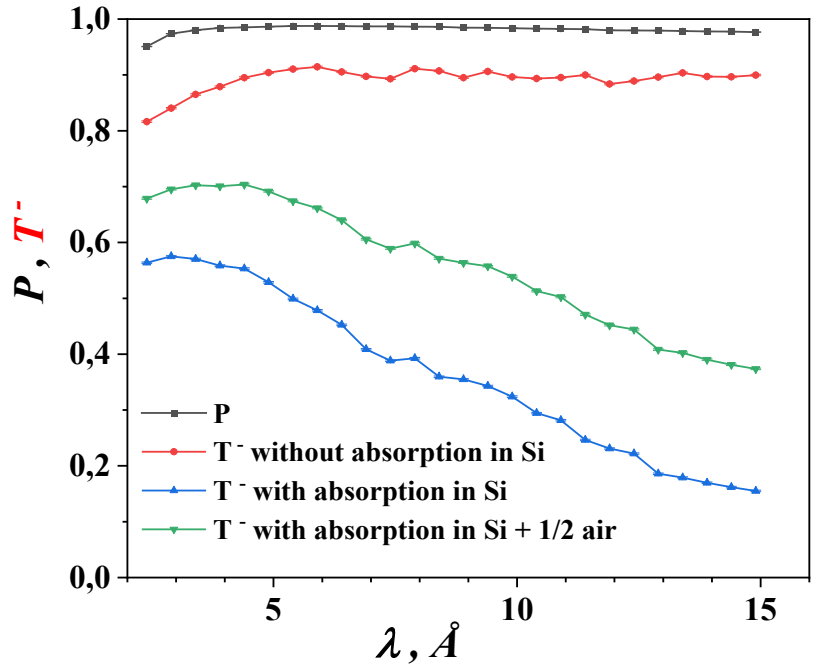

**Fig. 43.** Spectral dependences of the polarization $P$ and transmission $T^-$ of (-) spin component of the beam at the exit of the polarizer with optimal parameters for the wavelength range 2.4 – 15 Å for wafer lengths $L_1$ = 40 mm, $L_2$ = 45 mm, $L_3$ = 100 mm and for the total angular divergence of the input beam 0.9 degrees. Spectral transmission is represented by three curves: a curve without absorption and two curves with absorption: for a polarizer without air channels and for a polarizer with air channels.

For the wavelength range 0.9 – 2.4 Å, calculations of $P$ and $T^-$ for a polarizer with air channels were carried out using previously obtained optimal parameters for this wavelength range: the *CoFe/TiZr* parameter of the supermirror coating is $m$ = 2.5, the thickness of the silicon wafers and air channels is $d$ = 0.15 mm, the wafer lengths are $L_1$ = 44 mm, $L_2$ = 60 mm, $L_3$ = 100 mm. Input beam: angular divergence ± 0.20 degrees, beam width 30 mm.

Figure 44 shows the calculated spectral dependences of the polarization $P$ and the transmission coefficient $T^-$ for (-) spin component of the beam at the exit of a polarizer with air channels with optimal parameters for the wavelength range 0.9 – 2.4 Å. The spectral transmission coefficient is represented by three curves: a curve without absorption and two curves with absorption: for the polarizer without air channels and for a polarizer with air channels. As follows from the figure, the polarization of the output beam is high and exceeds the value of 0.95. The transmission coefficient $T^-$ is also high for a polarizer with air channels and its average value for the wavelength range $\lambda$ = 0.9 - 2.4 Å is 0.71, which exceeds the same value for a polarizer without air channels by 29%. Here, the absorption of the beam in aluminum was not taken into account.

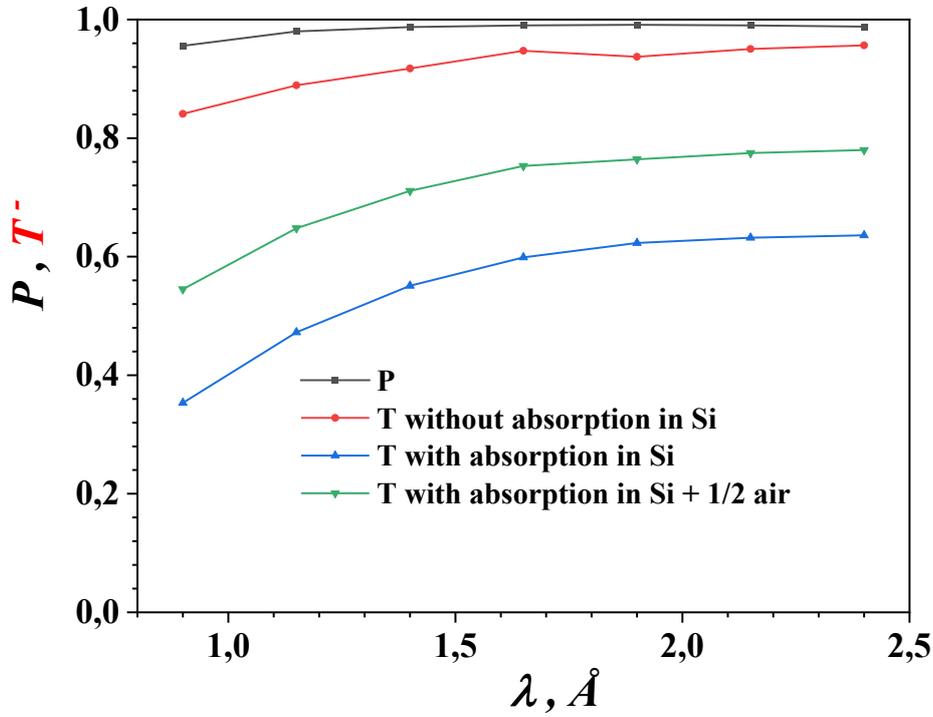

**Fig. 44.** Calculated spectral dependences of the polarization $P$ and the transmission coefficient $T^-$ for (-) spin component of the beam at the exit of the polarizer with optimal parameters for the wavelength range 0.9 – 2.4 Å for wafer lengths $L_1$ = 44 mm, $L_2$ = 60 mm, $L_3$ = 100 mm and for the total angular divergence of the input beam 0.4 degrees. Spectral transmission is represented by three curves: a curve without absorption and two curves with absorption: for a polarizer without air channels and for a polarizer with air channels.

Then, two polarizers with air gap channels and with optimal parameters for the wavelength range 0.9 - 2.4 Å were compared. The first polarizer is considered above with an electromagnetic system, the second polarizer consists of the same kink as the first, but its neutron guide is a Soller collimator (Fig. 9), the parameters of which are: $L_3$ = 100 mm and $d$ = 0.15 mm. In addition, a standard permanent magnetic element magnetic system is used for the second polarizer. In both cases, the input beam had a width of 30 mm and a beam angular divergence of ± 0.2 deg.

Figure 45 shows the calculated spectral dependences of the polarization P and the transmission coefficient $T^-$ for (-) spin component of the beam at the exit of the polarizers for both polarizers. The $T^-$ calculations were carried out taking into account the absorption: for the first polarizer in silicon 102 mm thick and aluminum 8 mm thick, for the second polarizer in silicon 102 mm thick. As follows from the figure, the polarization of the exit beam of the second polarizer is noticeably less than that of the first polarizer and its level is less than 0.95 in the wavelength range from 0.9 to 1.3 Å. The average $T^-$ level for the second polarizer is several times less than the same level of the first polarizer. *Appendix B* shows the dependence of the spectral transmission of a neutron flux through an aluminum layer with a thickness of 8 mm, taking into account absorption. This aluminum thickness is equal to the distance traveled by the beam through the turns of two polarizer coils with a wire cross section of 2 mm x 2 mm.

Figure 46 shows the spectral ratio of $T^-$ values for both polarizers. The average gain in intensity across the spectrum for the first polarizer is *4.4*.

Thus, the choice of the *SVAROG* polarizer is noticeably preferable in terms of luminosity compared to the polarizer with a *Soller* collimator for the wavelength range 0.9 – 2.4 Å. Calculations show that this conclusion is also valid when using *SVAROG* for the wavelength range of 2.4 – 15 Å.

As shown in *Appendix B*, the influence of imperfections in the polarizer design (shifts of the silicon wafers in the vertical and horizontal planes, deviation of the angles of the wafers in the assembly from the nominal values) slightly worsens its basic characteristics.

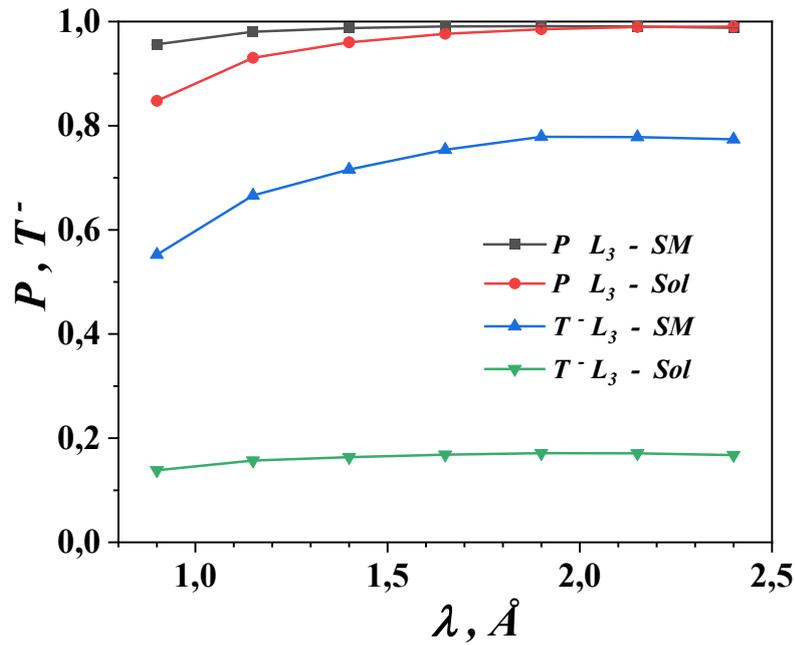

**Fig. 45.** Calculated spectral dependences of polarization $P$ and transmission coefficient $T^-$ for (-) spin component of the beam at the exit of two polarizers with optimal parameters for the wavelength range 0.9 – 2.4 Å, but with different coating of the wafers of neutron guide with length $L_3$. For the first polarizer with a supermirror coating (SM), for the second with an absorbing the coating is of the Soller type (Sol).

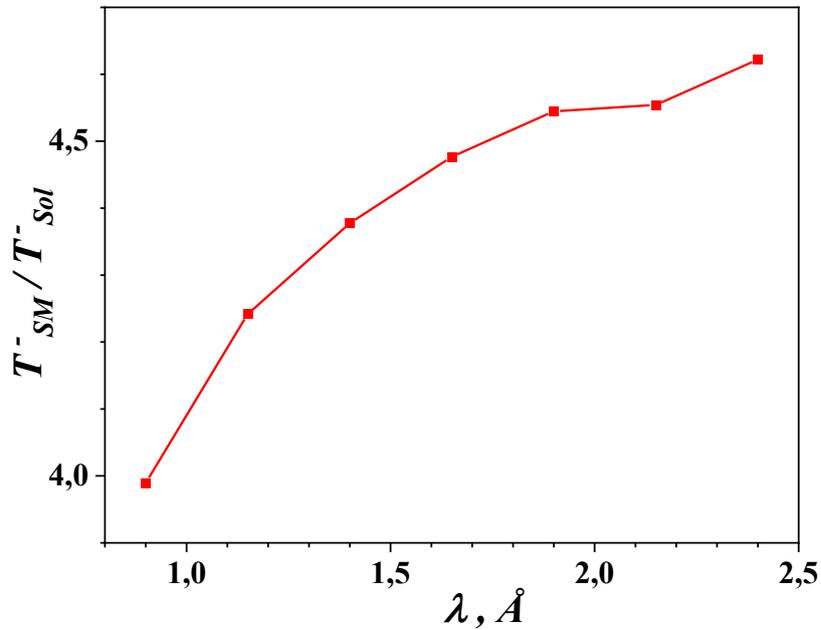

**Fig. 46.** The calculated gain curve in spectral transmission of the *SVAROG* polarizer compared to the kink polarizer with a Soller collimator.

## 7. Comparison of the *SVAROG* polarizer with its analogues.

The advantage of neutron transmission polarizers over *C*-bender polarizing reflectors is that the axes of the input and exit beams coincide.

**7.1. V-cavity.**

The V-cavity neutron transmission supermirror multichannel polarizer is widely used in international neutron centers [20-22]. Figure 47 shows a scheme of one V-cavity channel.

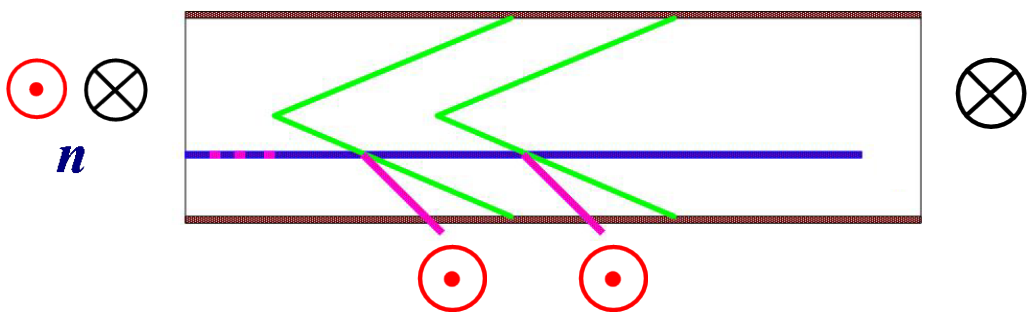

**Fig. 47.** A scheme of one V-cavity channel.

The principle of operation of this transmission polarizer and its features are described in [4]. The disadvantages of the V-cavity polarizer are: its large length of ~ 0.5 – 1.0 m even when using polarizing supermirror coatings with high values of $m$ ~ 5; the need to use several channels to work with beams of a sufficiently large area; dips in the spatial distribution of the beam intensity at the exit of the polarizer due to the separation glass plates between the channels of the polarizer, the high cost of supermirror coatings with high values of the supermirror parameter $m$ ~ 5.

## 7.2. Solid state multichannel benders.

Compact neutron transmission supermirror multichannel polarizers on neutron transparent substrates are known from the literature.

### 7.2.1. Solid state neutron polarizing multichannel S-shaped benders.

The supermirror solid state neutron polarizing multichannel $S$ - bender is also known [23, 24]. It is schematically shown in Fig. 48. It is a double-curved multichannel neutron guide on silicon wafers. The principle of operation of this transmission polarizer and its features are described in [4]. Disadvantages of the solid state $S$ - shaped bender: the beam passing through it significantly increases its angular divergence; the need to use supermirror coatings with a high value of the supermirror parameter $m$ > 3; difficulties in manufacturing such a large cross-section polarizer.

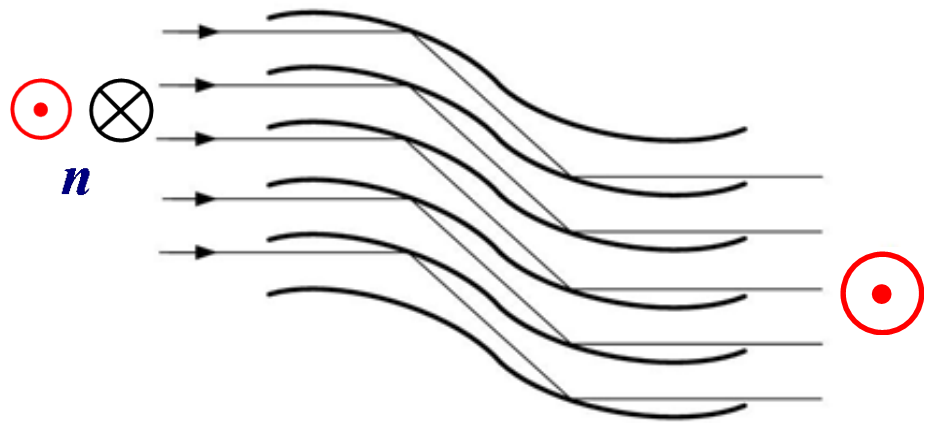

**Рис. 48.** Scheme of solid state neutron polarizing $S$ - shaped bender.

## 7.2.2. Solid state multichannel benders with collimators.

The supermirror solid state neutron polarizing multichannel bender with collimator is known [25]. This polarizer is schematically shown in Fig. 49. The polarizer consists of a bender on silicon wafers having a supermirror polarizing coating without an absorbing layer and a Soller collimator also on silicon wafers. The principle of operation of this transmission polarizer and its features are described in [4]. In [26], this polarizer is compared with the V-cavity.

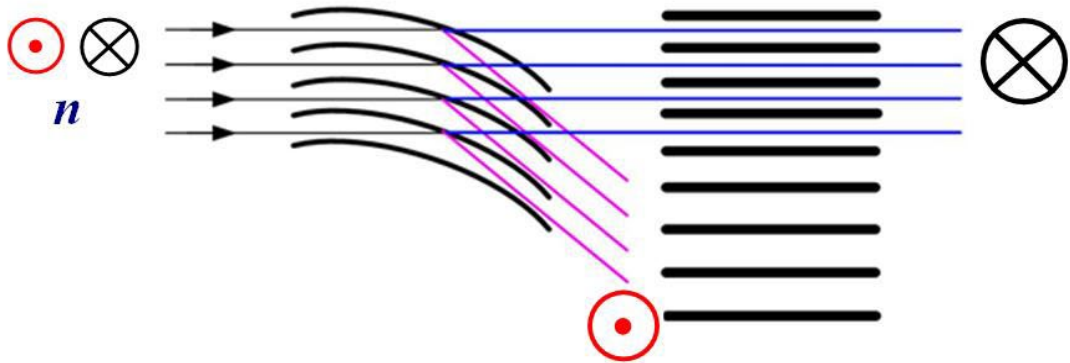

**Fig. 49.** The scheme of solid state neutron polarizing transmission multichannel bender with collimator.

In [27], various polarizing devices were compared for their applicability as a beam polarization analyzer in the MIEZE RESEDA facility. Based on this review, the choice was made in favor of solid state neutron polarizing multichannel bender with collimator. The paper also discussed the disadvantage of this polarizer. For neutrons with wavelengths near 5 Å, a noticeable spurious scattering was observed at the exit of the polarizer. The analysis revealed that this scattering occurs on bent bender silicon wafers.

In [28], various polarizing devices were compared for their applicability as a compact beam polarizer for a *FLEXX* triaxial cold neutron spectrometer with a wavelength range of $\lambda = 2 - 6$ Å. As a result, the choice, as in [27], was made in favor of solid state neutron polarizing multichannel bender with collimator. At the same time, the disadvantages of this polarizer were noted in [28]: low polarization at small and large wavelengths in the wavelength range under consideration, as well as low intensity of the transmitted beam compared to the reflective *C* - bender. In this work, an improved solid state neutron polarizing bender with collimator was proposed. The scheme of this interesting polarizer is shown in Fig. 50. The polarizing transmission bender is located between two straight polarizing remanent neutron guides (collimators) on silicon wafers. The collimators have a magnetization antiparallel to the guiding field, and the magnetization of the bender is parallel to the guiding field. Therefore, the walls of the collimators reflect well the neutrons of the (-) spin component and poorly reflect the neutrons of the (+) spin component, and the walls of the bender reflect well the neutrons of the (+) spin component and poorly reflect the neutrons of the (-) spin component. As shown in this work, the characteristics of the proposed polarizer (polarization and transmission) are noticeably better than those of a transmission bender with a collimator.

It can be noted that the disadvantages of the improved solid state neutron polarizing bender with collimator are: difficulties in manufacturing a large cross-section bender; the need to use supermirror coatings with a high parameter value *m* > 3 for polarizer elements; a problem when working with neutron wavelengths near 5 Å due to noticeable spurious beam scattering on bent silicon bender wafers; difficulty in organizing the necessary remanent states of the facility elements and eliminating possible depolarization of the beam during its travelling inside the polarizer. Although, one can try to solve the latter problem by using the concept of an electromagnetic system proposed and considered in this paper.

In the paper [28], it was also proposed, as a variant, to use saturating fields in polarizing collimators and the bender of an improved polarizer. It was also proposed to include two spin-flippers at the entrance and exit of the bender in this scheme. The disadvantage of this variant is that the distance between the polarizer elements will noticeably increase due to the need to ensure the correct operation of the two spin-flippers. In this case, the polarizer will no longer be compact. This question was discussed above, in section 3.3 of this paper.

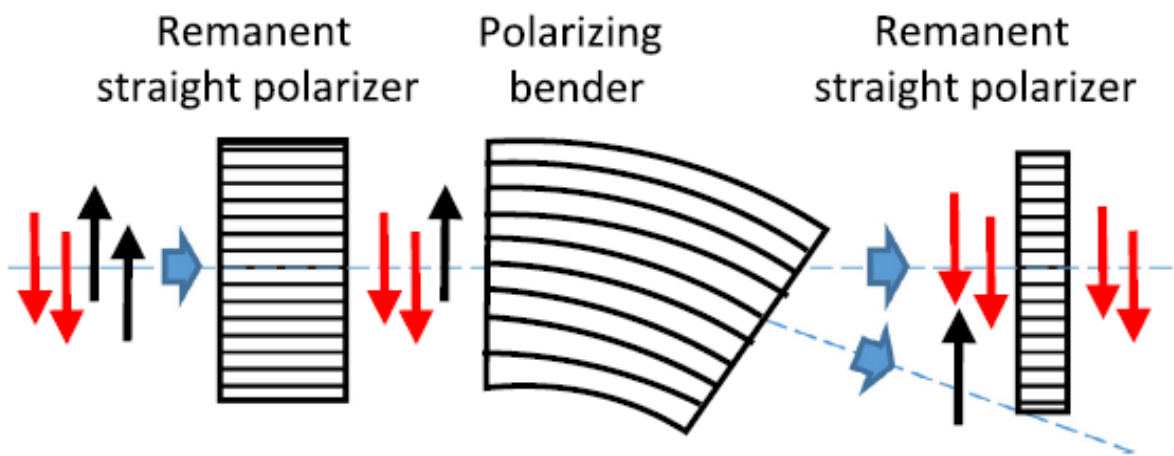

**Fig. 50.** The scheme of improved solid state neutron polarizing transmission multichannel bender with collimator [28].

# Conclusions:

I. A ***SVAROG*** - neutron transmission supermirror multichannel solid-state polarizer of a new type with an electromagnetic system is proposed.

II. The physical foundations of the creation of this polarizer are developed. Its parameters have been optimized.

III. Main features of the new polarizer:

1. The remanence. This makes it possible to work in magnetic fields with a field strength several times lower than the same value for a saturating field.

2. Compactness of the polarizer along the beam axis.

3. Work with beams having a significant cross-section (up to ~100x100 mm2) and a wide angular divergence.

4. The beam directions at the entrance to the polarizer and at its exit coincide.

5. The divergence of the output beam coincides with the divergence of the beam at the entrance to the polarizer.

6. The output beam has a uniform distribution in angle and coordinate.

7. The output beam has high values of polarization and intensity in a wide range of wavelengths $\lambda = 0.9 – 15$ Å.

8. It is shown that the influence of imperfections in the polarizer design slightly worsens its basic characteristics.

9. The *SVAROG* with air channels has a higher transmission compared to the option with only silicon wafers without air channels. In addition, this variant will require 2 times fewer plates, which will reduce the cost of the polarizer.

10. The luminosity of the *SVAROG* polarizer is several times higher than the luminosity of a kink polarizer with a Soller collimator.

IV. The main characteristics of the proposed polarizer are not worse than the main characteristics of the considered analogues, and in some characteristics they exceed them. For example, *SVAROG* surpasses V-cavity in compactness, simplicity of design, uniformity of the beam, and relative cheapness of the supermirror coating.

V. It is proposed to use *SVAROG* in experimental facilities, both at the *PIK* reactor and at other research reactors.

VI. It is planned to consider the polarizing supermirror *Fe/Si* as a coating for *SVAROG*, including the maximum value of the supermirror parameter $m \sim 5.5$.

## Statement of the author's contribution

**V.G. Syromyatnikov:** The idea of the *SVAROG* polarizer and its justification, Research planning, Calculation results processing and discussion, Writing the original text of the paper, Translating the text of the paper into English, Checking and Editing the paper.

**S.Yu. Semenikhin:** Preparation of the *Particle Raytracing* program and carrying out calculations according to this program, Carrying out calculations according to the *COMSOL Multiphysics 6.2* program, Discussion of calculation results, Discussion and Editing of the text of the paper.

**M.V. Lasitsa:** Calculations of spectral depolarization of a neutron beam using the *McStas* platform, Discussion of the calculation results, Discussion and Editing of the text of the paper.

## A statement of competing interests

The authors declare no conflict of interest. All authors have read and agreed to the published version of the manuscript.

## Acknowledgements

The authors consider it their pleasant duty to thank Dr. I.A. Zobkalo, Dr. A.N. Matveeva, A.P. Bulkin and K.Y. Terentyev for their interest in the work.

***Appendix A.*** Brief description of the *COMSOL Multiphysics 6.2* program. Calculation of the depolarization of a neutron beam passing through two solenoids of the *SVAROG* polarizer, depending on their parameters (wire cross-section and shape, insulation thickness, etc.).

A1. Brief description of the **COMSOL Multiphysics 6.2** program.

Calculations of the distribution of inter-turn magnetic fields depending on the parameters of the aluminum wire and the magnitude of the electric currents flowing through the coils were carried out using the *COMSOL Multiphysics 6.2* program [13]. This program is a universal software platform for numerical modeling of physical, chemical and biological processes. It is based on the finite element method (*FEM*) and allows solving complex problems related to electromagnetism, mechanics, heat transfer, acoustics, hydrodynamics and other fields. One of the key advantages of *COMSOL* is the possibility of multiphysical modeling, that is, the analysis of the interaction of several physical phenomena in one system. For example, it is possible to study the heating of a conductor under the action of an electric current, taking into account electromagnetic and thermal processes simultaneously. In stationary problems (Magnetostatics), *COMSOL* solves the equations of magnetostatics, taking into account the distribution of currents, permanent magnets, and magnetic materials. Examples of applications:
- analysis of the magnetic field of permanent magnets in sensors and electric motors;
- optimizing the shape of magnetic systems to achieve a uniform field;
- Simulation of magnetic shielding.

A2. Calculation of depolarization of a neutron beam passing through two polarizer solenoids *SVAROG* depends on their parameters (cross-section and shape of the wire, insulation thickness, etc.).

The calculated distributions of magnetic fields in the coils of the *SVAROG* polarizer, obtained in the *COMSOL* program, are shown in Fig. A1 and A2. The directions of electric currents in these coils are opposite. The arrows indicate the directions of the fields. The field in coils is 500 *Gs*. The cross-section of the wire in the coils is square with dimensions of 2 mm x 2 mm. The angular divergence of the beam in the horizontal plane is ± 0.25°.

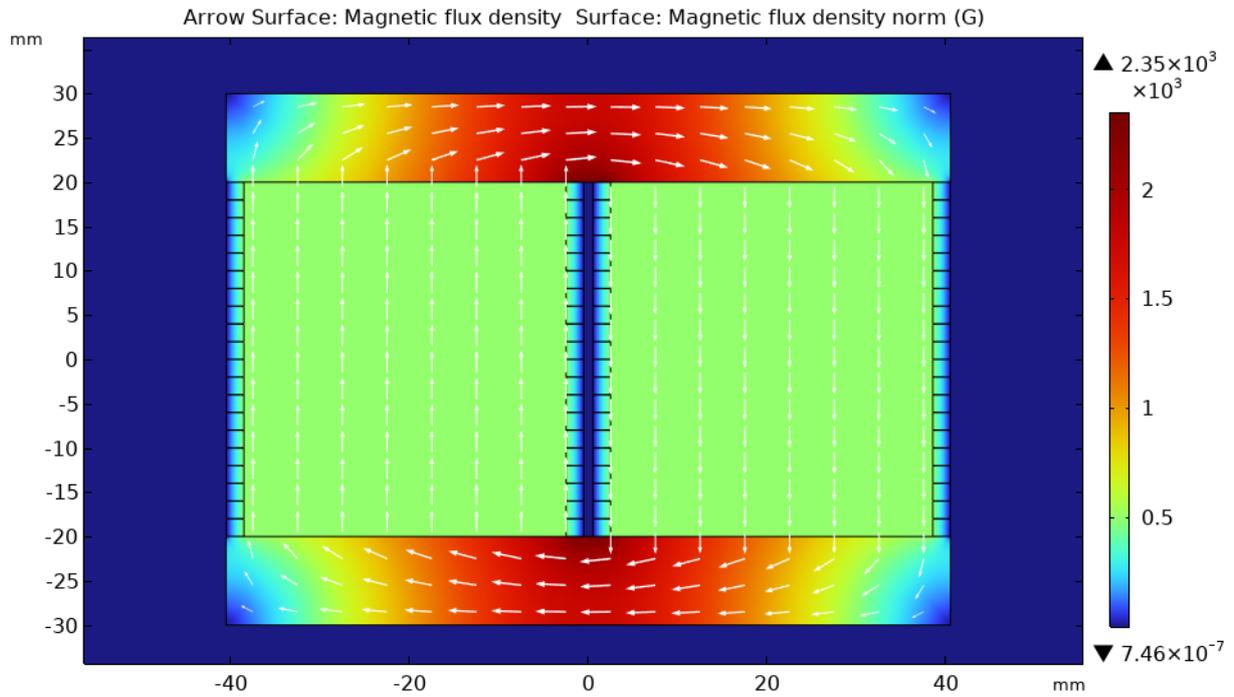

**Fig. A1.** The calculated distributions of magnetic fields in the coils of the *SVAROG* polarizer. The directions of electric currents in these coils are opposite. The arrows indicate the directions of the fields. The field in coils is 500 *Gs*. The cross-section of the wire in the coils is square with dimensions of 2 mm x 2 mm. The angular divergence of the beam in the horizontal plane is ± 0.25°.

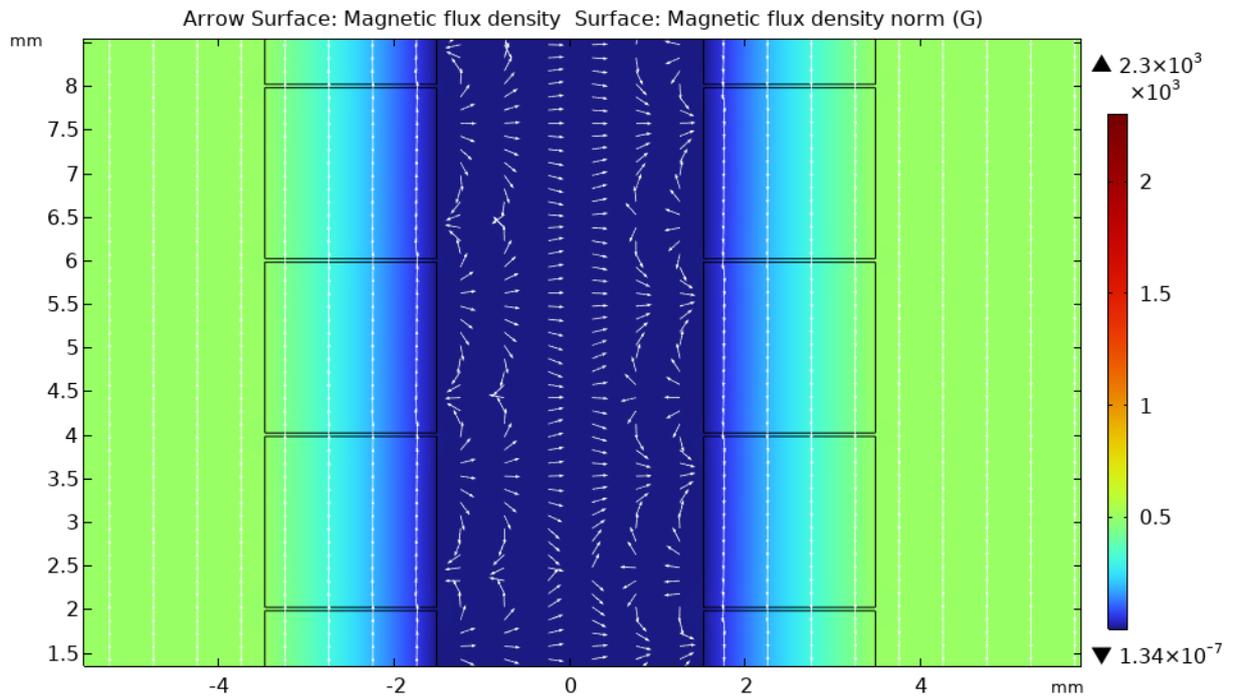

**Fig. A2.** The calculated distributions of magnetic fields in the coils within five turns. The directions of electric currents in these coils are opposite. The arrows indicate the directions of the fields. The field in coils is 500 *Gs*. The cross-section of the wire in the coils is square with dimensions of 2 mm x 2 mm. The angular divergence of the beam in the horizontal plane is ± 0.25°.

The effect of the parameters of the solenoid coils on the depolarization of the neutron beam passing through these coils was considered in detail in [14-17]. Aluminum plates of rectangular cross-section were used as turns in the coils.

Figures A3 – A9 show the results of calculations of spectral depolarization of the beam obtained for several parameters of the electromagnetic system of the *SVAROG* polarizer: two variants of the wire cross-sections of the solenoid coils, the thickness of the electrical insulation of the wire, and for several values of the distances between the coils.

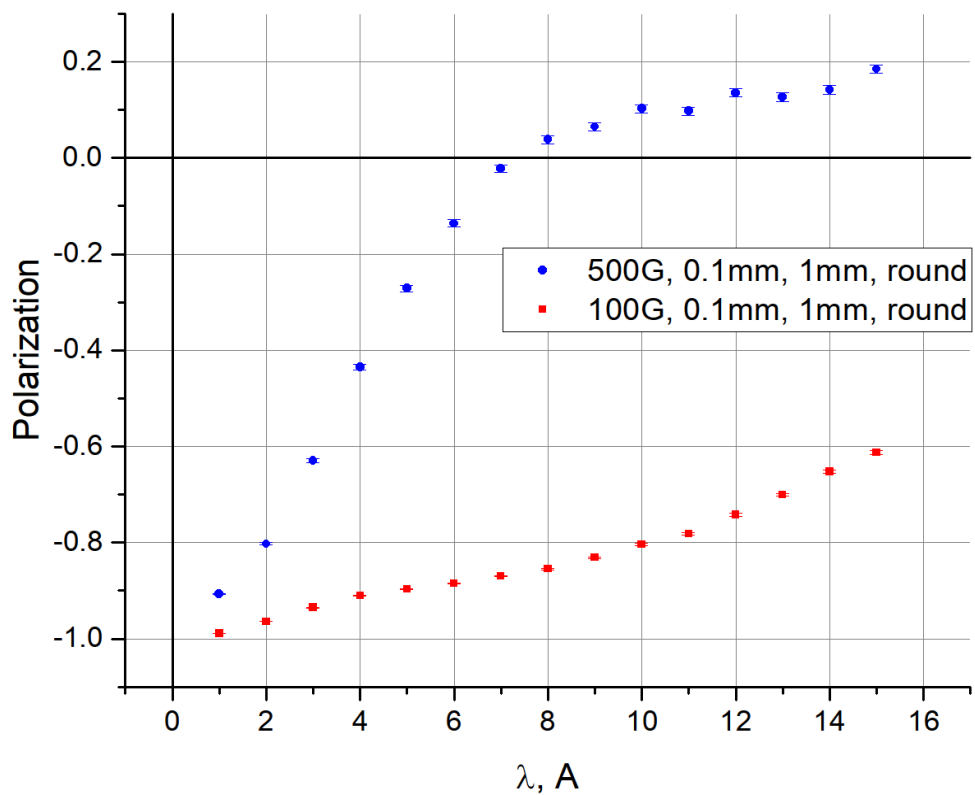

**Fig. A3.** Spectral depolarization of the beam for different fields in coils of solenoids. The thickness of electric insulation of the wire is 0.1mm. The distance between the coils is 1 mm. The wire cross-section is circular.

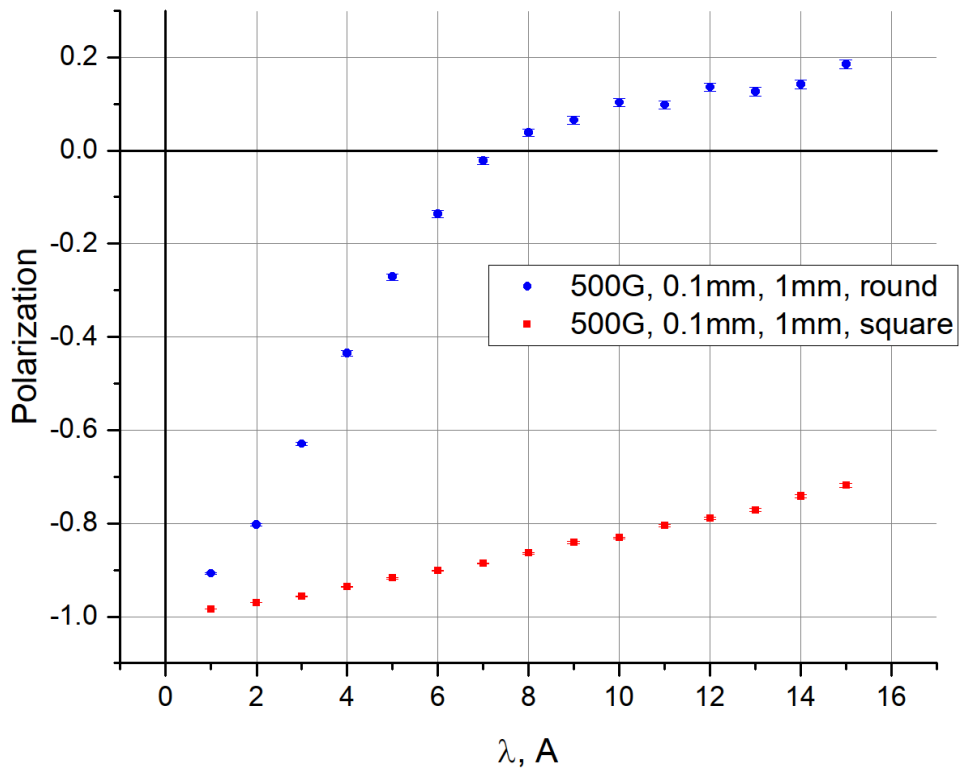

**Fig. A4.** Spectral beam depolarization for round and square wire sections of coils. The thickness of the electrical insulation of the wire is 0.1 mm. The distance between the coils is 1 mm.

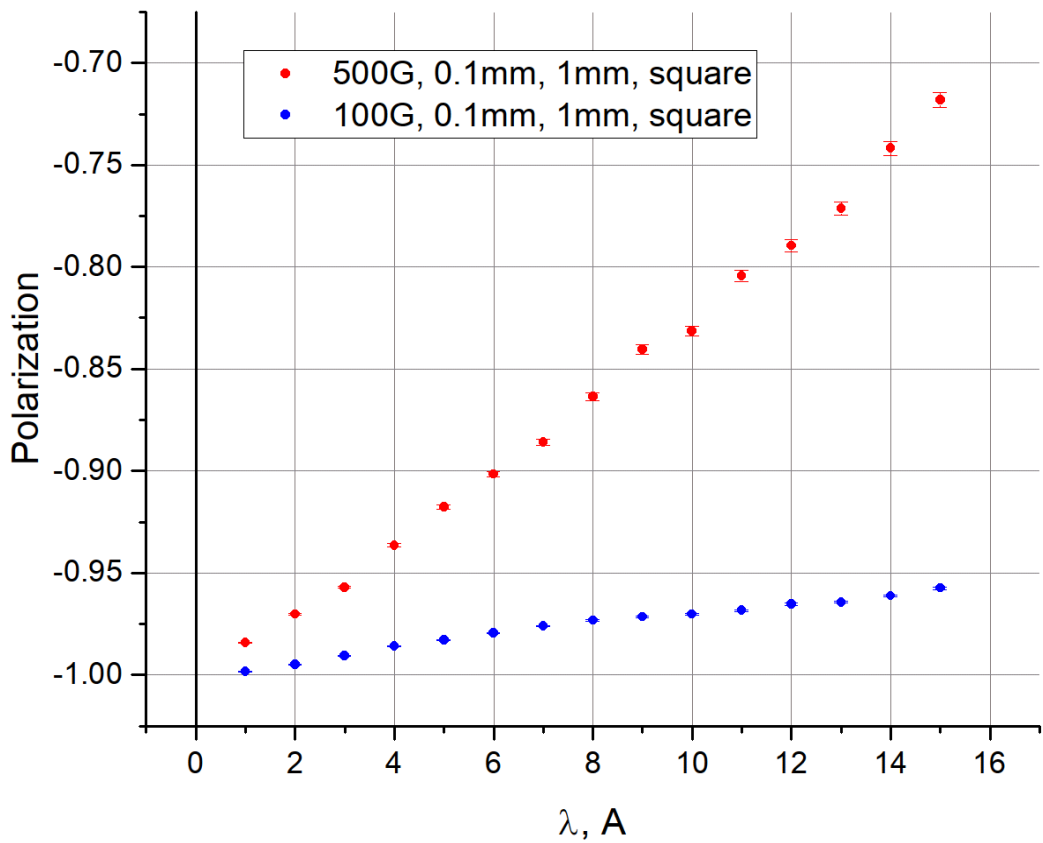

**Fig. A5.** Spectral beam depolarization for a 2 mm x 2 mm square wire section for different fields in solenoid coils. The thickness of the electrical insulation of the wire is 0.1 mm. The distance between the coils is 1 mm.

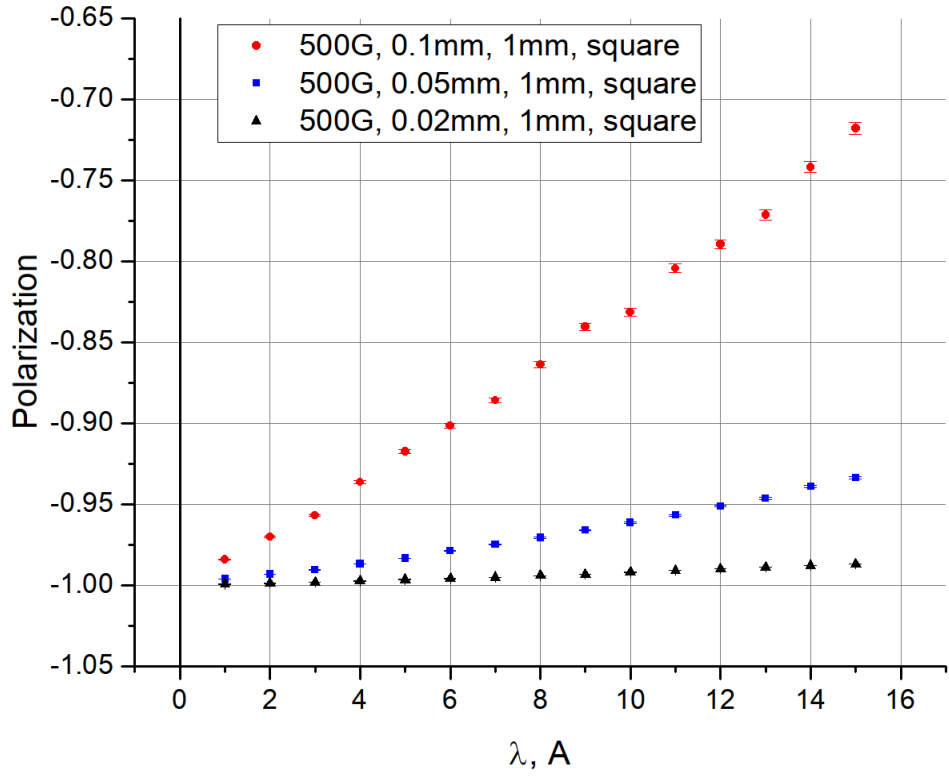

**Fig. A6.** Spectral beam depolarization for a square wire section of solenoid coils for wire electrical insulation thicknesses of 0.02, 0.05 and 0.1 mm. The distance between the coils is 1 mm.

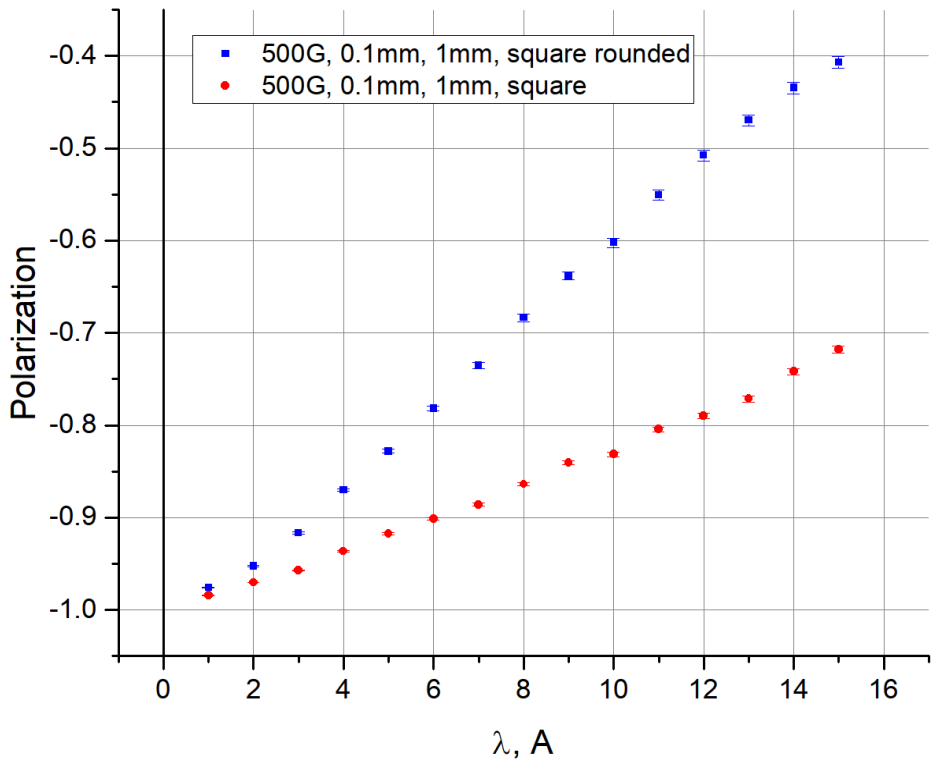

**Fig. A7.** Spectral beam depolarization for different wire sections: square and square with rounded edges (radius of rounding $r = 0.3$ mm). The thickness of the electrical insulation of the wire is 0.1 mm. The distance between the coils is 1 mm.

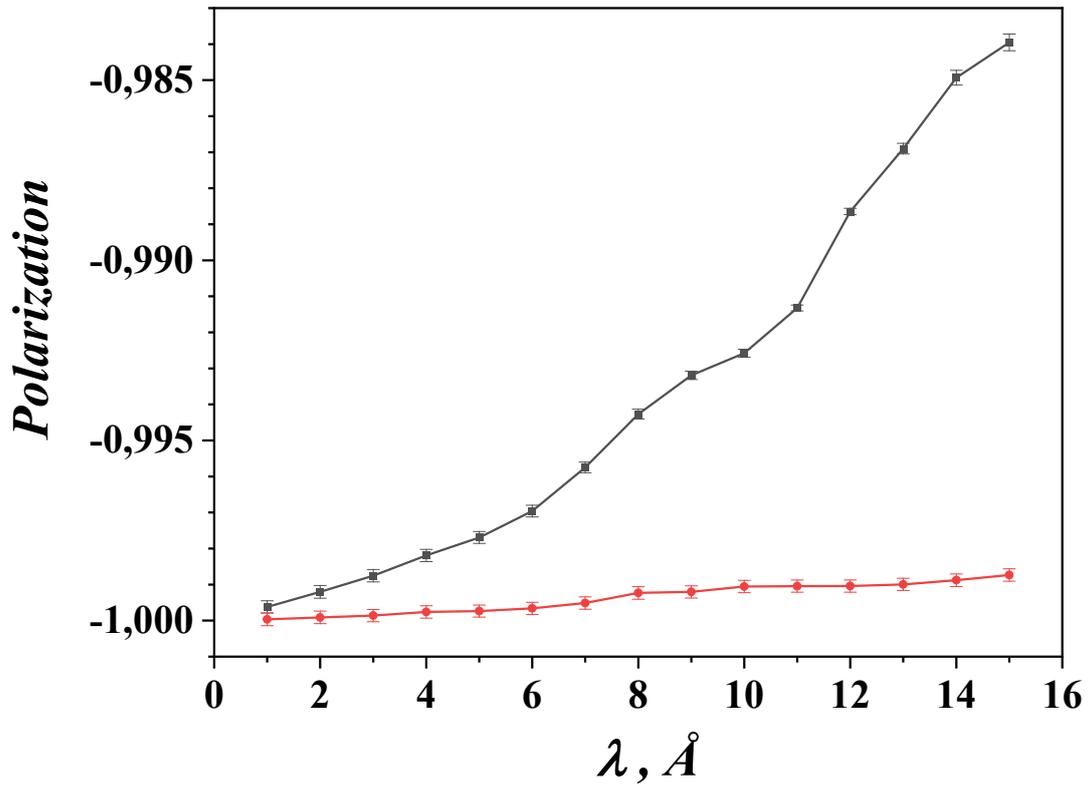

**Fig. A8.** Spectral depolarization of the beam for the distance between the coils of the solenoids of 6 mm. The fields in the coils are 500 *Gs* (black dots) and 100 *Gs* (red dots). The wire is square 2 mm x 2 mm. The electrical insulation thickness of the wire is 0.02 mm.

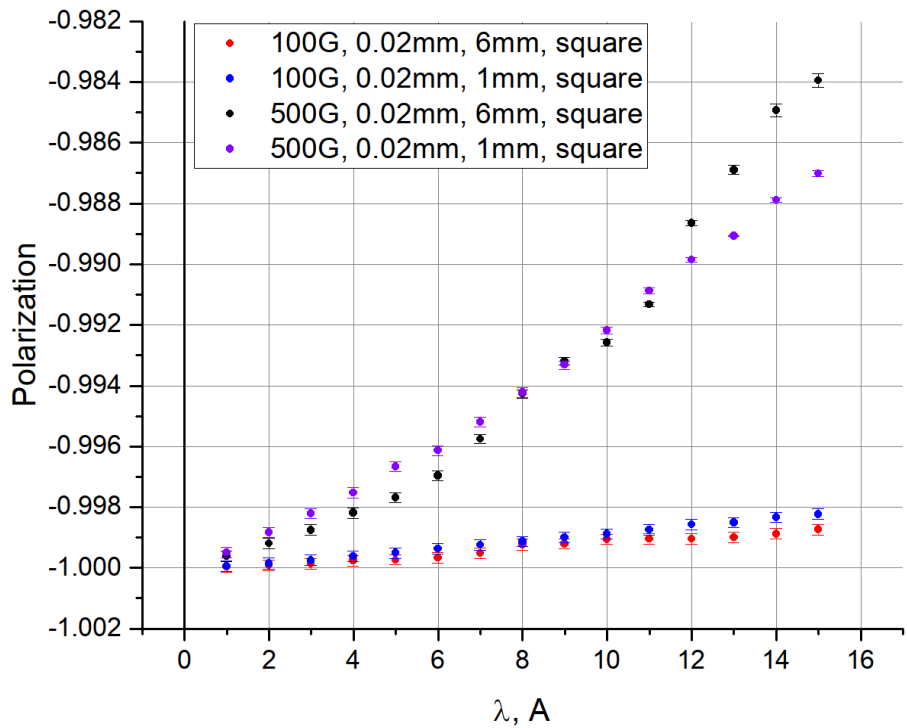

**Fig. A9.** Spectral depolarization of the beam for two distances between the coils of the solenoids: 1 mm and 6 mm. Fields in coils of 500 *Gs* and 100 *Gs*. The wire is square 2 mm x 2 mm. The electrical insulation thickness of the wire is 0.02 mm.

As a result of optimization, the following parameters of the electromagnetic polarizer system were obtained: aluminum wire with a square cross-section of 2 mm x 2 mm, the thickness of the electrical insulation of the wire is 0.02 mm, and the distance between the coils of solenoids kink and the neutron guide is 6 mm. Fields: in the 1st coil +85 *Gs*, in the 2nd coil -114 *Gs*, the guiding field between the coils, at the entrance to the polarizer and at the exit from the 2nd coil +15 *Gs*. Figure A10 shows the final results of calculating the spectral depolarization of a neutron beam passing through the coils of two electromagnets. As follows from Fig. A10, there is practically no spectral depolarization of the beam when passing through both coils with optimized parameters.

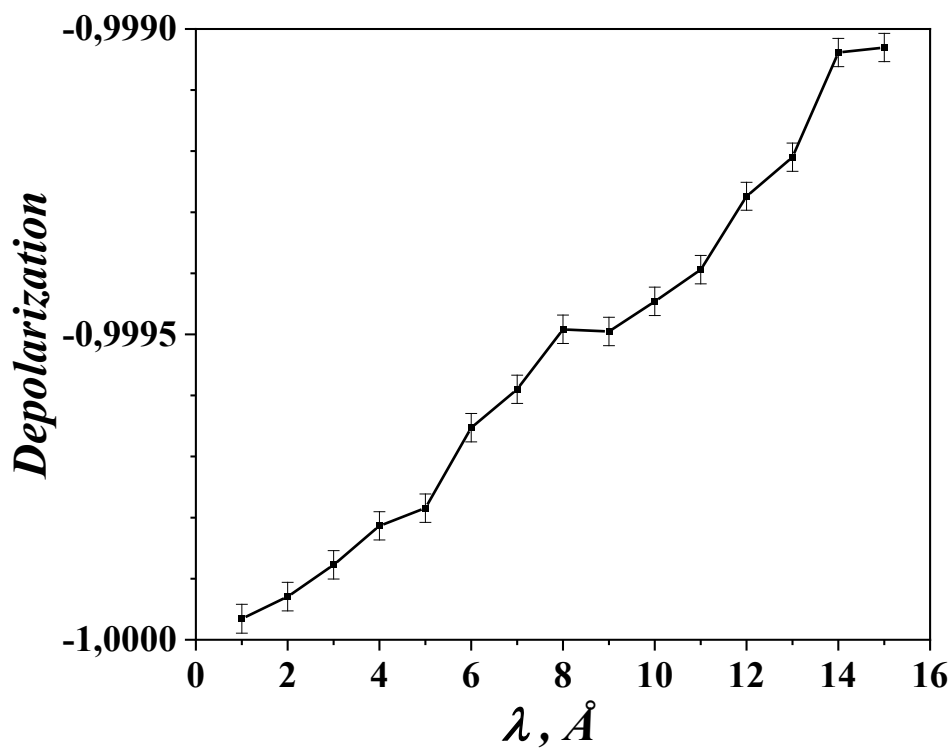

**Fig. A10.** The final results of the calculation of the spectral depolarization of a neutron beam passing through the coils of two electromagnets with optimized parameters.

*Appendix B.* Brief description of the *Particle Raytracing* program. Calculation of the effect of deviations from the correct geometry of the kink (shifts in the horizontal plane, shifts in the vertical plane, deviation of wafer angles in the assembly from nominal values) on the polarization and transmission of the beam at the exit of the polarizer. Accounting for beam losses in silicon and aluminum.

B1. Brief description of the *Particle Raytracing* program.

The *Particle Raytracing* program [18] is a new software tool for modeling neutron radiation and its interaction with various materials, which allows simulating neutron trajectories in complex facilities. The program is based on the Monte Carlo method, which provides statistical modeling of neutron scattering, absorption and reflection. The *Particle Raytracing* tool is written in the *Java* programming language. This software tool allows you to create your own tool modules based on ready-made tool classes, which can be combined to build virtual experimental facilities. The program supports parallel computing, which speeds up the processing of complex scenarios. Differences from the widely used similar *McStas* program: simplified *2D* modeling is used; real ray tracing is used, therefore, the reverse trajectory of particles in any of the facility tools is possible.

B2. The effect of the shift in the horizontal plane between the stacks of wafers in the kink.

Figure B1 shows a polarizer in which the stacks of wafers in the kink are shifted horizontally by an amount of $L_{12}$. In Fig. B2 shows the calculated dependences of the polarization $P$ and the transmission coefficient $T^-$ for (-) spin component of the beam at the exit from the polarizer on the distance $L_{12}$ between the kink stacks. Neutron absorption in silicon was not taken into account. Polarizer parameters: $m = 2.5$, $d = 0.3$ mm, $L_1 = L_2 = 35$ mm, $L_3 = 100$ mm. Beam parameters: $\lambda = 5$ Å, width 30 mm, angular divergence $\pm 0.45$ degrees.

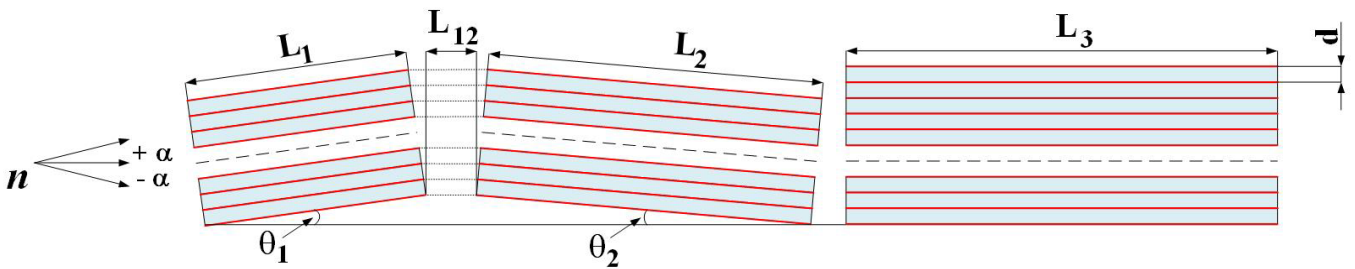

**Fig. B1.** The shift between the stacks of wafers in the kink in the horizontal plane by the amount of $L_{12}$.

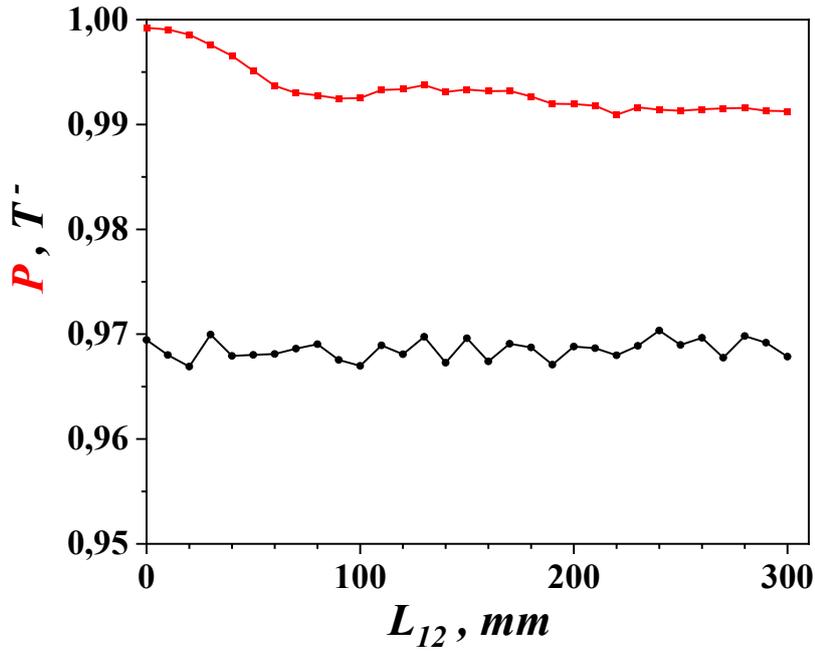

**Fig. B2.** Calculated dependences of the polarization $P$ and the transmission coefficient $T^-$ for (-) spin component of the beam at the exit from the polarizer on the shift between the kink stacks in the horizontal plane by the amount of $L_{12}$. Polarizer parameters: $m = 2.5$, $d = 0.3$ mm, $L_1 = L_2 = 35$ mm, $L_3 = 100$ mm. Beam parameters: $\lambda = 5$ Å, width 30 mm, angular divergence $\pm 0.45$ degrees.

As follows from Fig. B2, the transmission coefficient $T^-$ does not depend on the value of $L_{12}$ in the entire considered range of $L_{12}$ lengths. At the same time, the polarization of $P$ decreases slightly with increasing $L_{12}$, remaining very high.

B3. The effect of the shift in the vertical plane between the stacks of wafers in the kink.

In Fig. B3 shows a scheme of kink of a polarizer in which there is a shift of $\Delta y$ between the wafers in the stacks in the vertical plane in the direction perpendicular to the beam axis.

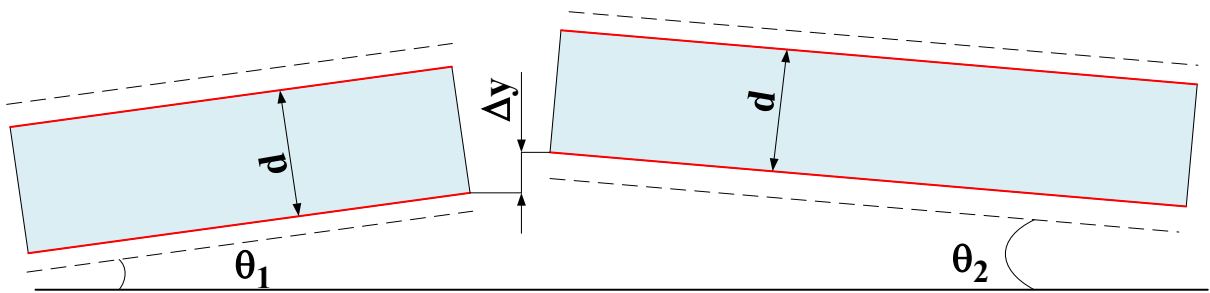

**Fig. B3.** The scheme of kink, in which there is a shift of $\Delta y$ between the wafers in the stacks in the vertical plane in the direction perpendicular to the beam axis.

Figure B4 shows the calculated dependences of the polarization $P$ and the transmission coefficient $T^-$ for (-) spin component of the beam at the exit of the polarizer on the magnitude of the shift $\Delta y$ between the wafers in the kink stacks in the vertical plane in the direction perpendicular to the beam axis. Neutron absorption in silicon was not taken into account. Polarizer parameters: $m = 2.0$, $L_1 = L_2 = 35$ mm, $L_3 = 100$ mm. Beam parameters: $\lambda = 5$ Å, width 30 mm, angular divergence $\pm 0.45$ degrees.

As follows from Fig. B4, the polarization $P$ and the transmission coefficient $T^-$ are weakly dependent on the magnitude of the shift $\Delta y$. When $\Delta y$ is changed by an amount equal to the thickness of the silicon wafer 0.3 mm, the values of $P$ and $T^-$ return to their initial values at $\Delta y = 0$.

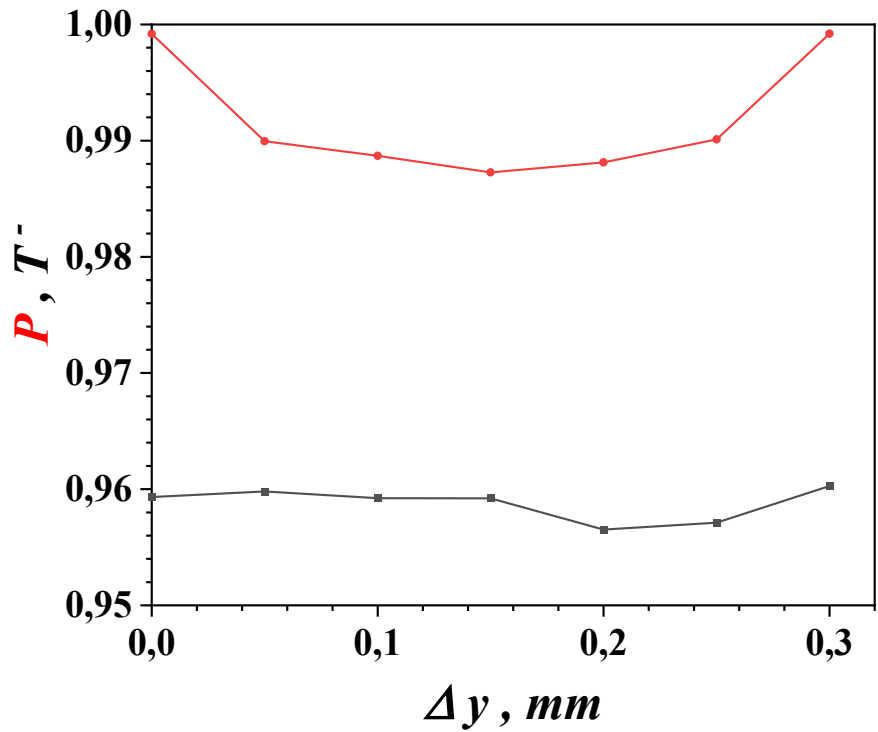

**Fig. B4.** Calculated dependences of the polarization $P$ and the transmission coefficient $T^-$ for (-) spin component of the beam at the exit from the polarizer on the magnitude of the shift $\Delta y$ between the wafers in the kink stacks in the vertical plane in the direction perpendicular to the beam axis. Neutron absorption in silicon was not taken into account. Polarizer parameters: $m = 2.0$, $d = 0.3$ mm, $L_1 = L_2 = 35$ mm, $L_3 = 100$ mm. Beam parameters: $\lambda = 5$ Å, width 30 mm, angular divergence $\pm 0.45$ degrees.

### B4. The effect of the deviation of the angles of the kink wafers from the nominal angle according to the Gaussian distribution

Figure B5 shows the calculated dependences of the spectral polarization $P$ of the beam at the exit of the polarizer for three values of the parameter $\sigma$, which characterizes the degree of deviation of the angles of the kink wafers from the nominal value $\theta_0$ according to the Gaussian distribution, as well as for two parameters of the supermirror *m* = 2.0 and 2.5. Polarizer parameters: *d* = 0.3 mm, $L_1 = L_2$ = 35 mm, $\theta_1 = \theta_2 = \theta_0$ = 0.49 degrees, $L_3$ = 100 mm. Beam parameters: width 30 mm, angular divergence ± 0.45 degrees. Three values of parameter *σ*: 1) *σ* = 0; 2) *σ* = 0.1 x $\theta_0$ = 0.049 degree; 3) *σ* = 0.15 x $\theta_0$ = 0.073 degree.

As follows from Fig. B5, the spectral polarization *P* weakly depends on the value of the parameter *σ*. For both values of the supermirror parameter *m*, the polarization decreases with increasing *σ* from 0 to 0.073 degrees within 1% and does not decrease below 0.95 for *m* = 2.0 and 0.97 for *m* = 2.5.

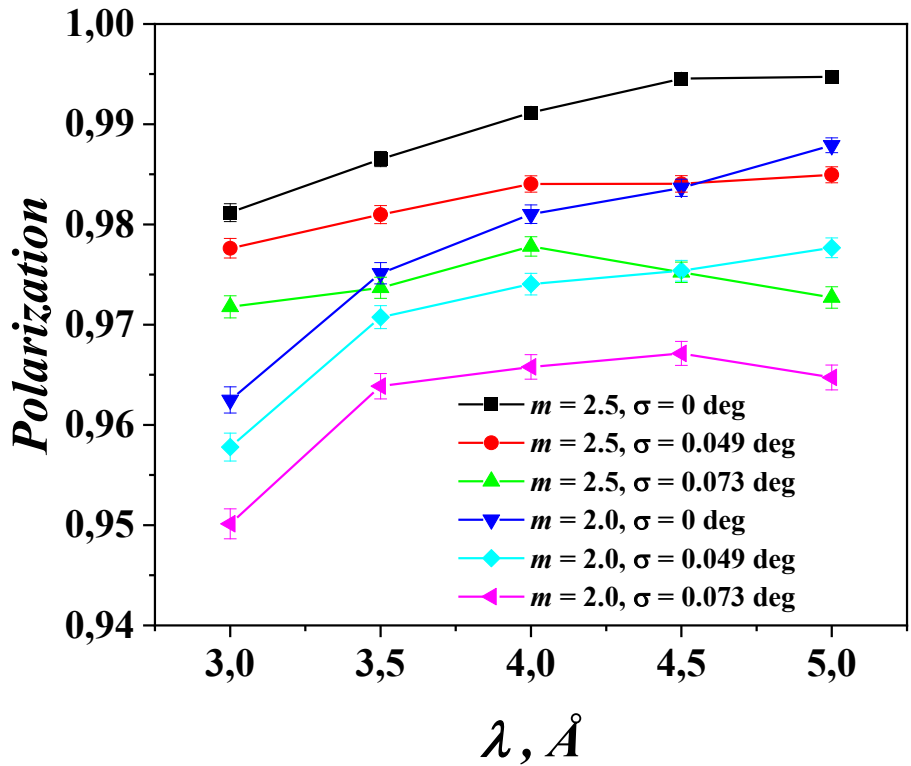

**Fig. B5.** Calculated dependences of the spectral polarization *P* of the beam at the exit of the polarizer for three values of the parameter *σ*, which characterizes the degree of deviation of the angles of the kink wafers from the nominal value $\theta_0$ according to the Gaussian distribution, as well as for two parameters of the supermirror *m* = 2.0 and 2.5. Polarizer parameters: *d* = 0.3 mm, $L_1 = L_2$ = 35 mm, $\theta_1 = \theta_2 = \theta_0$ = 0.49 degrees, $L_3$ = 100 mm. Beam parameters: width 30 mm, angular divergence ± 0.45 degrees.

B5. Absorption and scattering of neutrons in silicon and aluminum.

Figure B6 shows graphs of the dependence of the absorption and scattering cross-sections of neutrons in silicon on their energy, published in [19]. A curve for *single crystal (296 K) (open circles)* was used to account for neutron flux losses due to scattering and absorption during passage through silicon. After appropriate calculations, the dependence of the beam attenuation factor on the wavelength of the neutron and the effective path length of the neutron in silicon was obtained. This dependence, which takes into account the flow loss, was used in section No. 6 of this paper when calculating the transmission coefficient $T^-$ for (-) spin component of the beam as it passes through the polarizer.

Figure B7 shows the spectral dependence of the transmission of a neutron flux through an aluminum layer with a thickness of 8 mm, taking into account absorption. This aluminum thickness is equal to the distance traveled by the beam through the turns of two coils of polarizer with a wire cross-section of 2 mm x 2 mm.

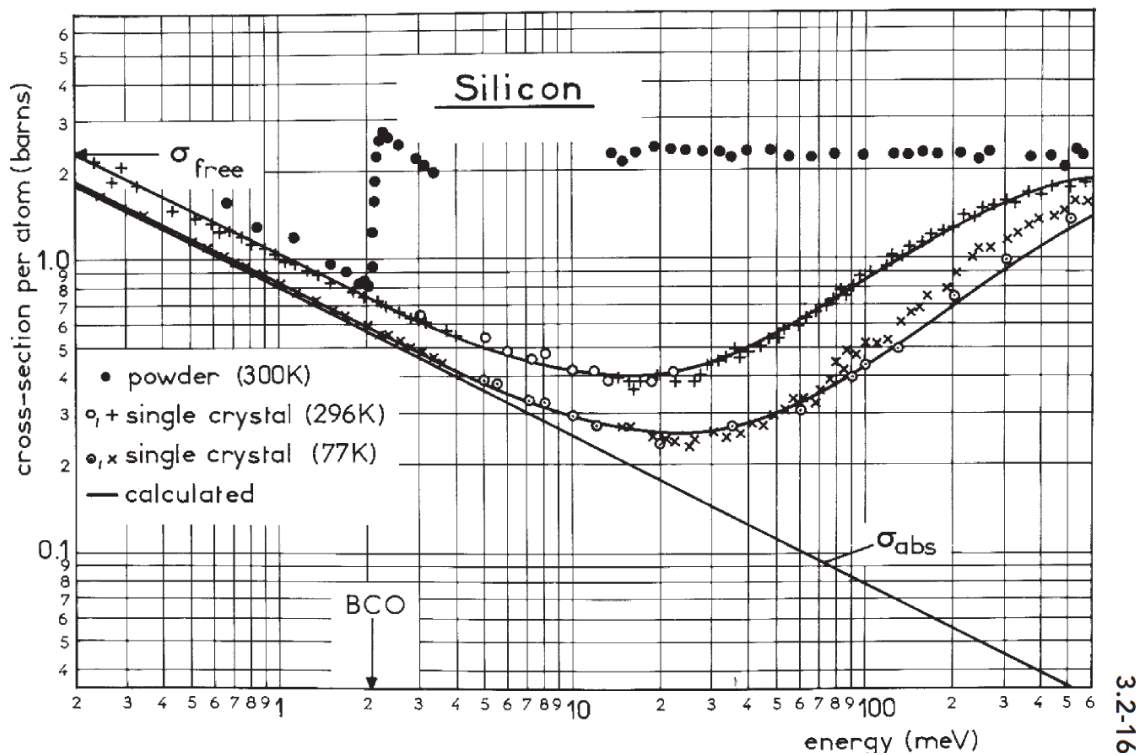

**Fig. B6.** Graphs of the dependence of the absorption and scattering cross sections of neutrons in silicon on their energy, published in [19]. A curve for *single crystal (296 K) (open circles)* was used to account for neutron flux losses due to scattering and absorption during passage through silicon.

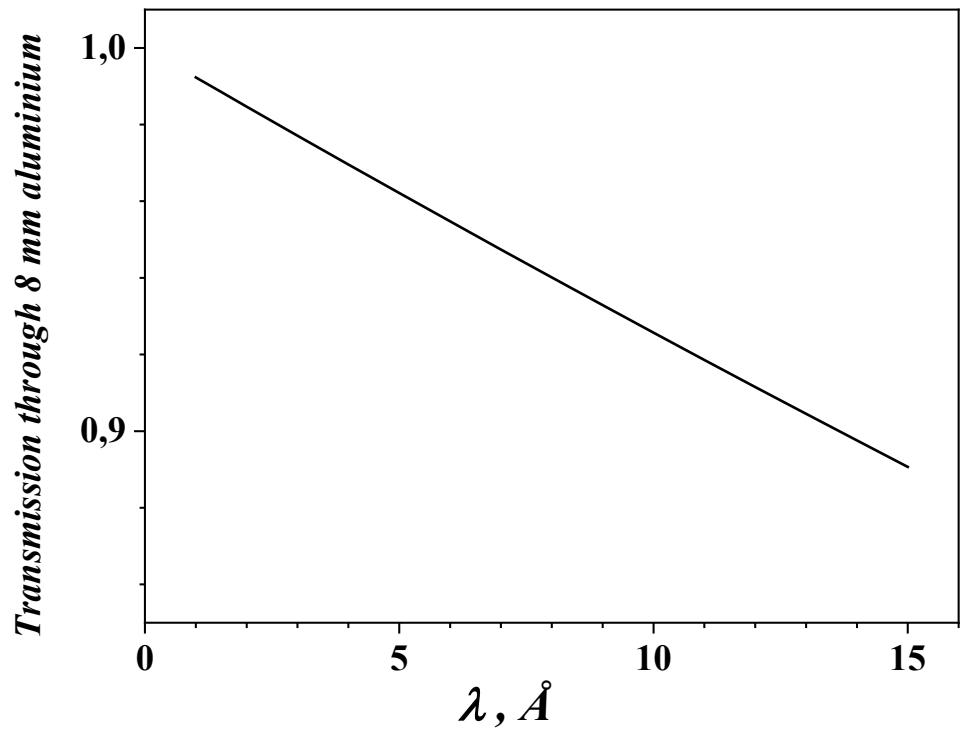

**Fig. B7.** Spectral dependence of the transmission of a neutron flux through an aluminum layer with a thickness of 8 mm, taking into account absorption.

# References:

[1]. V.G. Syromyatnikov. Russian patent N 2624633 for an invention. Priority from 21.06.2016.

[2]. V.G. Syromyatnikov, V.M. Pusenkov. Journal of Physics: Conf. Ser. **862** (2017) p. 012028.

[3]. V.G. Syromyatnikov. Russian patent N 2699760 for an invention. Priority from 13.12.2018.

[4]. V.G. Syromyatnikov. The paper in arXiv:1911.02936 [physics.ins-det], November 7, 2019 (28 pages).

[5]. V.G. Syromyatnikov. The paper in arXiv:2412.00223v1 [physics.ins-det] 29 Nov 2024 (21 pages).

[6]. V.G. Syromyatnikov, S.Yu. Semenikhin, M.V. Lasitsa. Proceedings of the Russian Conference on the Use of Neutron Scattering in Condensed Matter Research (RNIKS-2025) (September 29 - October 3, 2025, Tomsk, Russia), p. 53.

[7]. A.F. Schebetov, S.V. Metelev, B.G. Peskov, N.K. Pleshanov, V.M. Pusenkov, V.G. Syromyatnikov, V.A. Ul'yanov, W.H.Kraan, C.F. de Vroege, M.Th.Rekveldt. Physica B **335** (2003) p. 223.

[8]. Schebetov A.F., Pleshanov N.K., Syromyatnikov V.G. et al. Journal of Physical Society of Japan v. **65**, Suppl. A (1996) p. 195.

[9]. Pleshanov N.K., Peskov B.G., Pusenkov V.M., Syromyatnikov V.G., Schebetov A.F. Nuclear Instruments and Methods A **560** (2006) p. 464.

[10]. Boni, D. Clemens, M. Senthil Kumar, C. Pappas. Physica B **267-268** (1999) pp. 320–327.

[11]. J. Stahn, D. Clemens. J. Appl. Phys. **74** (2002) Suppl. S1532.

[12]. V.G. Syromyatnikov et al. Preprint Petersburg Nuclear Physics Institute, Gatchina, № 2619 (2005) 47 pages.

[13]. https://www.comsol.com/.

[14]. Thesis of Markus Bleuel. TU Munchen, 2003.

[15]. Thesis of Nikolas Arend. TU Munchen, 2007.

[16]. R. Georgii, N. Arend, P. Boni, D. Lamago, S. Muhlbauer, and C. Pfleiderer. Neutron News, v. **18**, Number 2 (2007) pp. 25-28.

[17]. R. Georgii, G. Brandl, N. Arend, W. Häußler, A. Tischendorf, C. Pfleiderer, P. Böni, and J. Lal. Appl. Phys. Lett. **98**, issue 7 (2011) p. 073505.

[18]. S.Yu. Semenikhin. *The Particle Raytracing* program for calculating the transmission coefficients of a neutron beam through neutron-optical systems. It will be published.

[19]. Neutron Data Booklet ILL (Editors: Albert-Jose Dianoux ILL (Grenoble), Gerry Lander ITU (Karlsruhe)), p. 3.2-16, April 2002. Freund A.K. Nuclear Instruments and Methods **213**, issues 2-3 (1983) pp. 495-501.

[20]. F. Mezei. Proc. SPIE **983** (1988) p. 10.

[21]. T. Keller, T. Krist, A. Danzig, U. Keiderling, F. Mezei, A. Weidemann. Nuclear Instruments and Methods A **451** (2000) p. 451.

[22]. https://www.swissneutronics.ch/

[23]. Krist Th., Rucker F., Brandl G., Georgii R. Nuclear Instruments and Methods A **698** (2013) p. 94.

[24]. A. Stunault, K.H. Andersen, S, Roux, T. Bigault, K. Ben-Saidane, H.M. Ronnow. Physica B **385-386** (2006) pp. 1152-1154.

[25]. Krist Th., Peters J., Shimizu H.M., Suzuki J., Oku T. Physica B **356** (2005) p. 197.

[26]. J. Repper, W. Haussler, A. Ostermann, L. Kredler, A. Chacon and P. Boni. Journal of Physics: Conference Series **340** (2012) p. 012036

[27]. Jochum, J.K., Cooper, J.F.K., Vogl, L.M., Link, P., Soltwedel, O., Böni, P., Pfleiderer, C., Franz, C. Quantum Beam Science **6** (2022) p. 26.

[28]. J. Xu, M. Atterving, M. Skoulatos, A. Ostermann, R. Georgii, T. Keller, and P. Böni. Nuclear Instruments and Methods A **1031** (2022) p. 166526.